\documentclass[twocolumn,prd,,preprintnumbers,nofootinbib,aps,floats,floatfix,amsmath,amssymb,secnumarabic]{revtex4} %

\usepackage{graphicx}
\usepackage[usenames,dvipsnames]{color}
\usepackage{amsmath,amssymb}
\usepackage{slashed}
\usepackage{verbatim}
\usepackage[colorlinks,citecolor=blue]{hyperref}
\usepackage[utf8]{inputenc}

\newcommand{\be}{\begin{equation}}
\newcommand{\ee}{\end{equation}}
\newcommand{\ba}{\begin{array}}
\newcommand{\ea}{\end{array}}
\newcommand{\bea}{\begin{eqnarray}}
\newcommand{\eea}{\end{eqnarray}}
\newcommand{\sss}{\scriptscriptstyle}

\newcommand{\HF}{{\sss \rm FH}}

\newcommand{\nn}{\nonumber}

\newcommand{\tGamma}{{\tilde \Gamma}}

\def\sfrac#1#2{{\textstyle\frac{#1}{#2}}}
\def\lsim{\mathrel{\raise.3ex\hbox{$<$\kern-.75em\lower1ex\hbox{$\sim$}}}}
\def\gsim{\mathrel{\raise.3ex\hbox{$>$\kern-.75em\lower1ex\hbox{$\sim$}}}}

\def\R{{\rm R}}
\def\L{{\rm L}}

\def\tL{{t_{\rm L}}}
\def\bL{{b_{\rm L}}}
\def\tR{{t_{\rm R}}}

\begin{document}
%

\title{Electroweak baryogenesis at high wall velocities}

\author{James M.\ Cline}
\email{jcline@physics.mcgill.ca}
\affiliation{Department of Physics, McGill University, 
             3600 Rue University, Montr\'eal, Qu\'ebec, Canada H3A 2T8}
\author{Kimmo Kainulainen}
\email{kimmo.kainulainen@jyu.fi}
\affiliation{Department of Physics, P.O.Box 35 (YFL), 
             FIN-40014 University of Jyv\"askyl\"a, Finland}
\affiliation{Helsinki Institute of Physics, P.O. Box 64,
             FIN-00014 University of Helsinki, Finland}
\affiliation{Theoretical Physics Department, CERN, 1211 Geneva 23, Switzerland}

\begin{abstract} 
It is widely believed that electroweak baryogenesis should be
suppressed in strong phase transitions with fast-moving bubble walls,
but this effect has never been quantitatively studied. We rederive
fluid equations describing transport of particle asymmetries near the
bubble wall without making the small-wall-velocity approximation. We show that
the suppression of the baryon asymmetry is a smooth function of the
wall speed and that there is no special behavior when crossing
the sound speed barrier. Electroweak baryogenesis can thus be
efficient also with strong detonations, generically associated with
models with observably large gravitational waves. We also make a
systematic and critical comparison of our improved  transport
equations to another one commonly used in the literature, based on the
VEV-insertion formalism.
\end{abstract}
\preprint{CERN-TH-2019-227}
\pacs{ }
\maketitle

%
\section{Introduction}
%

Electroweak symmetry exhibits a phase transition in the early universe, that is known to be a smooth crossover in the standard model (SM) 
16
 \cite{Kajantie:1996mn}, but could become first order if new physics beyond the SM couples significantly to the Higgs boson.  A strongly first order electroweak phase transition (EWPT) is one of the necessary requirements for electroweak baryogenesis (EWBG) \cite{Bochkarev:1990fx,Cohen:1990py,Cohen:1990it,Turok:1990zg}, and it could also be a source of gravity waves that might be observed at LISA.  

There is a perception that EWBG and observable gravity waves would tend to be mutually exclusive however, since the latter require very strong phase transitions, which lead to fast-moving bubbles, with wall velocity $v_w\!\sim \!1$.  This makes it difficult for  particle
asymmetries to diffuse efficiently in front of the wall and bias sphalerons to create the baryon asymmetry.  There may be some tension between the two effects, but until recently there have been few quantitative studies including transport of the particle asymmetries \cite{Vaskonen:2016yiu,Dorsch:2016nrg,Bian:2017wfv,Akula:2017yfr,Huang:2018aja,Grzadkowski:2018nbc}. Most works have focused on the coexistence of observable gravitational waves with the sphaleron washout
condition $v/T\gtrsim 1$
\cite{Huang:2016cjm,Artymowski:2016tme,Hashino:2016xoj,Chao:2017vrq,Beniwal:2017eik,Kurup:2017dzf,Baldes:2017rcu,Angelescu:2018dkk,Beniwal:2018hyi,Ahriche:2018rao,Athron:2019teq}, 
without taking into account the problem of  reduced particle transport near the wall.  

The theoretical deficit is in large part due to the fact that transport equations for the particle asymmetries have been derived using the approximation $v_w\ll 1$, making it impossible to reliably predict the baryon asymmetry at large $v_w$.   Since it was believed that $v_w\lesssim 0.1$ in the SM \cite{Moore:1995ua,Moore:1995si} and the minimal supersymmetric standard model (MSSM) \cite{John:2000zq}, the small-$v_w$ approximation seemed adequate at the time.  In recent years however, there has been increased interest in two-step phase transitions involving a scalar singlet field \cite{Espinosa:2011ax}, which is able to generate stronger phase transitions with typically higher $v_w$. Such transitions can more easily satisfy the sphaleron washout constraint, and in addition can be a strong source of gravitational waves.   It is therefore timely to revisit the transport equations relevant for EWBG and try to extend their applicability to higher $v_w$. We will show in particular that nothing special happens when the wall speed crosses the sound barrier, and that the baryon asymmetry only vanishes smoothly in the extreme limit $v_w\to 1$.\footnote{Throughout this work, $v_w$ is taken as a proxy for the relative speed between the bubble wall and the plasma in front of it.  In realistic solutions of the fluid equations near the wall, it can happen that this relative velocity, which is the relevant quantity for diffusive transport, differs from the wall velocity as measured with respect to the plasma at infinite distance.}

This work has two main goals. The first is to update the fluid equations for the semiclassical force mechanism~\cite{Joyce:1994zt,Cline:2000nw,Kainulainen:2001cn,Kainulainen:2002th,Prokopec:2003pj,Prokopec:2004ic} to arbitrary wall velocities. This is strongly motivated because the currently existing formulation~\cite{Fromme:2006wx} breaks down for wall velocities exceeding the sound speed. Our second purpose is to perform a quantitative comparison between the semiclassical method and the competing ``VEV-insertion'' approximation~\cite{Riotto:1995hh,Riotto:1997vy}.
These two approaches agree that the particle densities contributing to the baryon asymmetry are determined by (quantum) Boltzmann equations, but it remains controversial what precise form they should take. 

The semiclassical method is designed to be valid when the de Broglie wavelength of the
particles, of order the inverse temperature $T^{-1}$, is smaller than the typical width of a
bubble wall $L_w$. The interactions of particles with the wall can then be treated as coming from a semiclassical force, that can be derived using the WKB approximation~\cite{Cline:1997vk,Cline:2000nw,Cline:2001rk}, or from the closed-time-path (CTP) formalism of thermal field theory~\cite{Kainulainen:2001cn,Kainulainen:2002th,Prokopec:2003pj,Prokopec:2004ic}.
In the semiclassical approach the CP-violating force appears at the level of the Boltzmann equations.
It is straightforward to approximate them by a set of moment equations with source terms induced by the force, that can be determined systematically in an expansion in powers of  $(L_w T)^{-1}$ (though the subleading corrections have not been computed).  

The VEV-insertion method is also derived starting from the CTP formalism. Here quantum Boltzmann equations are manipulated to yield their classical counterparts at the level of integrated particle densities. In this approach the source term is not easy to extract and one must make a rather drastic approximation, expanding a two-point function to leading order in the spatially varying Higgs field VEV $v(z)$~\cite{Riotto:1995hh,Riotto:1997vy}. This is known as the VEV-insertion approximation. It can be regarded as an expansion in powers of $v(z)/T$, which cannot be very small inside the bubble if the phase transition is sufficiently strong to avoid washout. It is hoped that since $v(z)$ is somewhat smaller inside the bubble wall, this can still be a reasonable approximation. But if that is the case, it must be capturing quite different physics from the WKB approach, since the two formalisms cannot be obviously reconciled, and in general they make quite different predictions. 

For example, EWBG in the MSSM was analyzed using both formalisms~\cite{Cline:1997vk,Cline:2000kb,Cline:2000nw,Carena:1996wj,Carena:1997gx}, with the VEV-insertion method giving significantly larger estimates for the asymmetry. However a systematic study of the differences between the two methods is lacking in the literature, in particular in comparing their predictions as a function of parameters characterizing the bubble wall.
We will provide such a comparison in this work, for a prototypical model of CP violation in the wall. As was the case for studies of EWBG in the MSSM, we will
demonstrate a large discrepancy between the predictions of the two methods.

We start in section~\ref{sect:vw} by arguing that the transport equations should not suffer from any sort of critical behvior for bubble walls that move near the speed of sound, but should rather only do so as $v_w\to 1$. In section~\ref{sec:improved} we review the derivation of the WKB transport equations and the origin of their $v_w$ dependence. We point out an inconsistency in the approximations used in ref.~\cite{Fromme:2006wx} (hereafter denoted FH06), and remedy it by a more careful evaluation of the coefficient functions for general values of $v_w$. In section~\ref{sec:model} we introduce our fiducial model and in section~\ref{sec:comparisonFH06} we compare the predictions of the FH06 equations and our improved fluid equations in the semiclassical approach.  In sect.~\ref{sec:fcomp} we quantitatively compare the semiclassical approach to the VEV insertion framework.  Conclusions are given in sect.~\ref{sec:dictionary}.

%
\section{Relevance of wall velocity}
\label{sect:vw}
%

The basic idea for reducing the full Boltzmann equations to a set of coupled  first order fluid equations for the chemical potential and velocity perturbation, in the context of electroweak baryogenesis, was  set out in ref.\ \cite{Joyce:1994zt}.  The method was elaborated for the MSSM in ref.\  \cite{Cline:2000nw} and for general two-Higgs doublet models in ref.~\cite{Fromme:2006wx}. In principle, one can always ``integrate out'' the velocity perturbation and convert the coupled system into a single second order diffusion equation for the chemical potentials, as was done in ref.\  \cite{Cline:2000nw}.  However this is complicated by the fact that any particle that couples strongly to the Higgs boson (as required to source electroweak baryogenesis) has a mass that varies within the  bubble wall, and therefore the coefficients in the diffusion equation are functions of $z$, the distance transverse to the wall.  If one makes the crude approximation of ignoring baryon violation  by sphalerons inside the bubble and taking the masses to vanish outside, then the $z$-dependence goes away and Green's functions techniques can be used to solve the diffusion equations.  However this effectively approximates the wall as being very thin, which is inconsistent with the semiclassical expansion underlying the whole fluid approach.  For quantitative results, one should numerically solve the coupled equations keeping track of the full $z$-dependence of the coefficients.

An important contribution of FH06 was to calculate all of the coefficient functions $K_i^\HF(x)$, where $x=m(z)/T$, appearing in the fluid equations.  In their approach, these functions are independent of the wall velocity $v_w$, which was achieved by expanding to leading order in $v_w$.  Therefore one could question to what extent this formalism can be accurate for walls with large
wall velocities.

\begin{figure}[t]
\hspace{-0.4cm}
\centerline{\includegraphics[width=0.95\hsize]{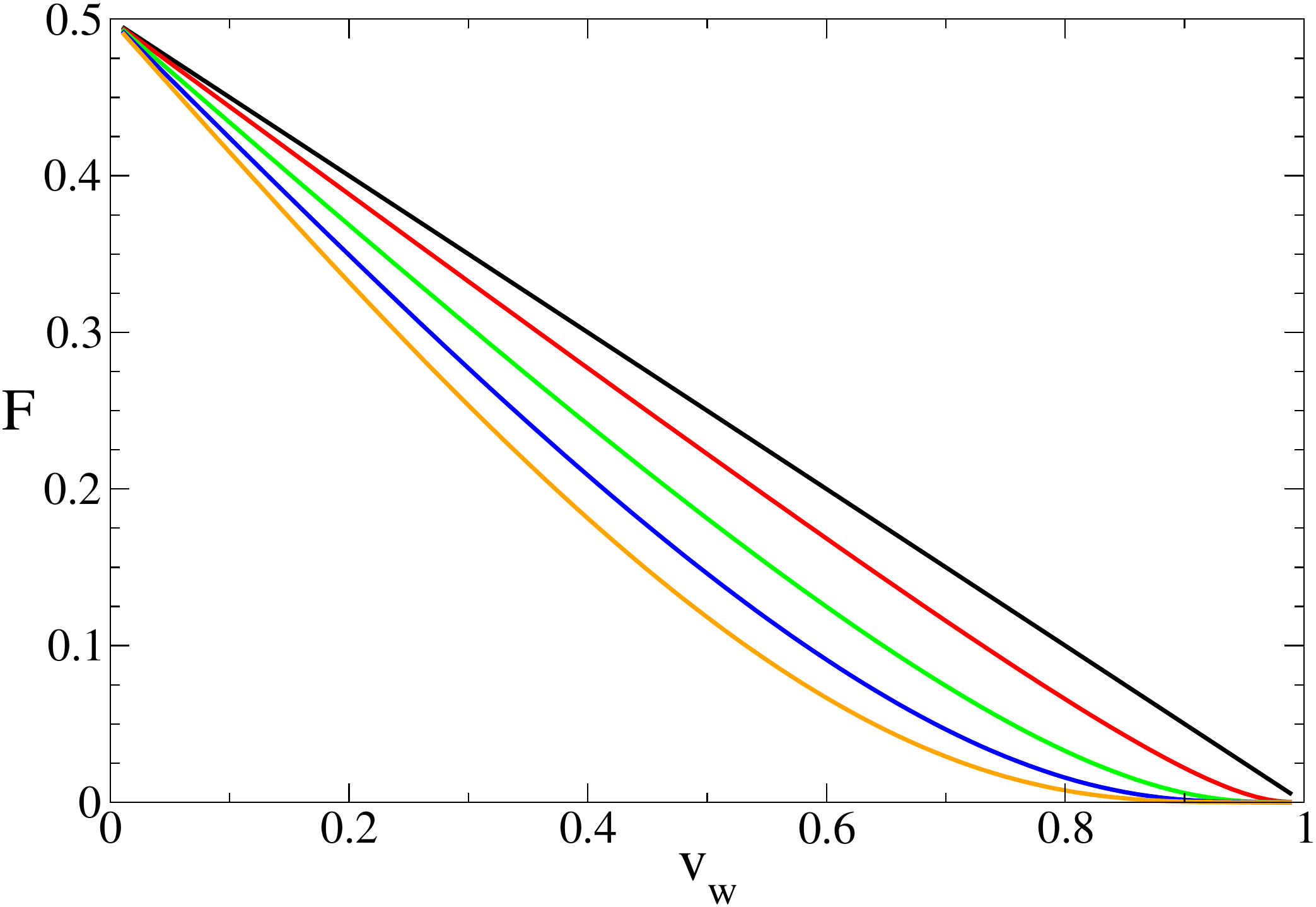}}
\caption{Fraction of plasma particles that can stay ahead of a bubble wall moving
at speed $v_w$.  Different curves are for fermions with $m/T = 0,1,2,3,4$ (top to bottom).}
\label{frac}
\end{figure}

It is expected that diffusion lengths should diminish as $v_w$ gets large. One can make a simple estimate to quantify this statement, by asking what fraction $F$ of particles in the plasma are moving faster than the wall, in its direction of motion. We would expect that only of order this fraction is able to contribute to the diffusion tail for the CP asymmetry in front of the wall. It is straightforward to show\footnote{by first doing the angular integral over $\cos\theta = p_z/p$} that
\be
  {\rm F} = \frac{\sfrac12\int_{\gamma_w v_w m}^\infty {\rm d}p\, p^2 (1 - v_w E/p)/(e^{\beta E} + 1)}
  {\int_{0}^\infty {\rm d}p\, p^2 /(e^{\beta E} + 1)}
\ee
for a massive fermion.  In fig.~\ref{frac} we plot ${\rm F}$ versus $v_w$ for particles with 
increasing values of $m/T$.  It is clear that nothing dramatic happens near the sound speed
$v_s\cong 1/\sqrt{3}$; instead $v_w=1$ is the only critical speed.

%
\section{Derivation of transport equations}
\label{sec:improved}
%

The Boltzmann equation acting on the unperturbed distribution functions can be written in the wall frame as
\be
\left(v_g\partial_z + F\partial_{p_z}\right) f = {\cal C}[f] \,.
\label{L2}
\ee
For a fermion with a CP-violating complex mass term $\hat m(z) = m(z)e^{i\gamma^5\theta(z)}$~\cite{Cline:2000nw,Kainulainen:2002th},
\bea
  v_g &=& \frac{p_z}{E_w}
\label{eq:vgF_1a}
\\
  F   &=& -\frac{(m^2)'}{2E_w} + ss_{k_0} \frac{(m^2\theta')'}{2 E_w E_{wz}},
\label{eq:vgF_1b}
\eea
where $^\prime$ denotes $\partial_z$.
Here $E_{wz}^2 = E_w^2 - {\bf p}_{||}^2$ and $E_w$ is the conserved wall frame energy. $s_{k_0} =1$ for particles and $-1$ for antiparticles, and $s=\pm 1$ for the states that are the eigenstates of the spin $s$ in $z$-direction in the frame where the momentum of the state parallel to the wall ${\bf p}_{||}$ vanishes. For the wall frame helicity eigenstates one should replace~\cite{Cline:2017qpe}\footnote{We correct a typo in equation (E3) of~\cite{Cline:2017qpe}, by replacing $|p_z|\rightarrow p_z$.}
\be
s \rightarrow s_h  = h\gamma_{||}\frac{p_z}{|{\bf p}|} \equiv h s_{\rm p} \,,
\label{eq:shel}
\ee
where $h=\pm1$ is the helicity and $\gamma_{||} = E_w/E_{wz}$ is the Lorentz boost for going to the frame where ${\bf p}_{||}=0$. In practice the difference between the two spin bases is small~\cite{Cline:2017qpe}. In particular in the massless limit $s_h = h\,{\rm sign}(p_z)$. Equation~\eqref{eq:vgF_1a} is actually the definition of the physical momentum $p_z$ from the group velocity determined by the WKB dispersion relation~\cite{Cline:2000nw,Kainulainen:2002th}. It is convenient to write this relation in a form that defines $E_w$ in terms of the physical momentum:
\be
E_w \approx E - s_hs_{k_0} \frac{m^2\theta^\prime}{2EE_z} \equiv E + s_hs_{k_0}\Delta E,
\label{eq:Ew}
\ee
where $E \equiv \sqrt{{\bf p}^2 + m^2}$ and $E_z \equiv \sqrt{p_z^2 + m^2}$. Using these variables, eqs.~(\ref{eq:vgF_1a}-\ref{eq:vgF_1b}) become.:
\bea
  v_g &=& \frac{p_z}{E} + s_hs_{k_0}  \frac{m^2\theta'}{2 E^2 E_z} 
\label{eq:vgF_2a}
\\
  F &=& -\frac{(m^2)'}{2E} + s_hs_{k_0} \left( \frac{(m^2\theta')'}{2 E E_z} 
        -\frac{m^2 (m^2)'\theta'}{4 E^3 E_z}\right).\;
\label{eq:vgF_2b}
\eea
Eqs.\ (\ref{eq:Ew}-\ref{eq:vgF_2b}) agree with those derived in~\cite{Fromme:2006wx} when one sets $s_h \rightarrow s$.

For bosons there is no CP-violating semiclassical force at this order in the gradient expansion~\cite{Cline:2000nw}. However the CP-even kinetic force remains, and so all equations are valid for bosonic degrees of freedom if one simply sets $s_{k_0}=0$ everywhere.

The starting point for deriving fluid equations from the Boltzmann equation is to expand 
particle distribution functions around the equilibrium distribution. Because the kinetic momentum $p_z$ is conserved in collisions, the expansion in the rest frame of the bubble wall 
looks like
\be
  f = {1\over e^{\beta[\gamma_w(E_w + v_w p_z)-\mu]} \pm 1} + \delta f,
\label{fexp}
\ee
where $\gamma_w = 1/\sqrt{1-v_w^2}$, Here $\mu$ is a pseudochemical potential that defines the particle asymmetry and $\delta f$ is an extra term whose specific form should be left unspecified,\footnote{Unlike the perturbation in the chemical potential, whose algebraic form is enforced by fast elastic scattering processes, the form of the velocity perturbation is not predictable~\cite{Cline:2000nw}. Thus, assuming a specific ansatz for its shape in momentum space can lead to unphysical behavior, in particular at large $v_w$.} except for stipulating that
\be
  \int {\rm d}^3p\, \delta\! f = 0\,.
\label{dfeqs_1}
\ee
This condition is just the definition of $\mu$; it  ensures that $\delta\! f$ does not affect the local particle density. 

%
\subsection{Classification by CP-parity}
%

Next observe that the semiclassical force in~\eqref{eq:vgF_2b} contains two distinct pieces: the first, CP-even term, is equal for particles and antiparticles while the second, CP-odd term, is opposite for particles and antiparticles. The CP-even term is of first order in gradients while the CP-odd term is of second order. Because of this hierarchy, one can solve the CP-even and CP-odd equations separately. To this end we introduce the definitions
\bea
\mu &\equiv& \mu_e + s_{k_0}\mu_o 
\nn\\
\delta\! f &\equiv& \delta\! f_e + s_{k_0}\delta\! f_o 
\eea
Using these together with  eq.\ \eqref{eq:Ew} we can write eq.\ \eqref{fexp} as
\be
f \approx f_{0w} + \Delta f_e + s_{k_0}\Delta f_o \,,
\label{eq:fexpansion}
\ee
where, expanding to leading consistent order in both CP-even and CP-odd quantities,
\bea
\Delta f_e &=&  -\mu_e f^\prime_{0w} + \delta f_e \nn\\
\Delta f_o &=&  (-\mu_o + s_h\gamma_w\Delta E) f^\prime_{0w} \nn\\
&& \phantom{mmn} - s_h\gamma_w\Delta E f^{\prime\prime}_{0w}\mu_e  + \delta f_o,
\eea
where prime denotes ${\rm d}/{\rm d}(\gamma_wE)$ and 
\be
f_{0w} = \frac{1}{e^{\beta[\gamma_w(E_w + v_w p_z)]} \pm 1} \,.
\ee

The expansion~\eqref{eq:fexpansion} is also necessary for bosons. Even though bosonic equations do not have direct CP-violating sources at the order to which we are working, they can inherit CP-violating perturbations from their interactions with fermions.

To derive the CP-even equation we drop the CP-odd parts proportional to $s_{k_0}$ in the expansion~\eqref{eq:fexpansion} and in eqs.~(\ref{eq:vgF_2a}-\ref{eq:vgF_2b}) for the group velocity and the semiclassical force. After this the Boltzmann equation \eqref{L2} immediately becomes
\be
L[\mu_e,\delta f_e] = {\cal S}_e + \delta {\cal C}_e,
\label{eq:CPeven}
\ee
where the Liouville operator is defined as
\bea
L[\mu,\delta f] \! &\equiv& \! -\frac{p_z}{E}f_{0w}^\prime \,\partial_z\mu + v_w\gamma_w\frac{(m^2)^\prime}{2E}f^{\prime\prime}_{0w} \mu \nn\\
&& + \frac{p_z}{E} \partial_z \delta f - \frac{(m^2)^\prime}{2E}\partial_{p_z}\delta f 
\label{eq:Liouville}
\eea
and the CP-even source term is:
\be
{\cal S}_e = v_w \gamma_w \frac{(m^2)^\prime}{2E} f_{0w}^\prime .
\label{eq:CPeven_source}
\ee
The collision term for the CP-even perturbation $\delta {\cal C}_e$ is model dependent and we do not specify it further until sect.\ \ref{sec:model}. The CP-even equations (\ref{eq:CPeven}-\ref{eq:CPeven_source}) are valid both for bosons and fermions, and are helicity independent, unlike their CP-odd counterparts.

%
\subsection{CP-odd equation}
%

In the CP-odd sector we must account for the helicity. Because the relevant physical quantity for EWBG is the left-handed chiral asymmetry in front of the wall, one often concentrates only on the negative helicity sector,\footnote{This is reasonable when masses vanish in front of the wall. If this is not the case, one should compute the asymmetry in the positive helicity sector as well, and project out the left chiral asymmetry from both helicity contributions. Note that while the Liouville terms are identical for both helicities, the sources are equal and opposite. The collision terms are also helicity dependent.} but to be
general we keep the full helicity dependence. Projecting out the CP-odd part of the Boltzmann equation~\eqref{L2} requires some work, but the final result is analogous to eq.\ \eqref{eq:CPeven} up to source and collision terms: 
\be
L[\mu_o,\delta f_o] = {\cal S}_o + \delta {\cal C}_o \,,\phantom{Ha}
\label{eq:CPodd}
\ee
where the CP-odd source term is
\bea
{\cal S}_{oh} 
   &=&- v_w \gamma_w hs_{\rm p} \frac{(m^2\theta')'}{2 E E_z} f'_{v_w} \nn\\
   && + v_w\gamma_w hs_p \frac{m^2(m^2)^\prime\theta^\prime}{4 E^2 E_z}
	\left({f'_{v_w}\over E} - \gamma_w f''_{v_w}\right)
\label{eq:CPodd_source}
\eea
and the collision integral $\delta C_o$ is again model dependent, which
we will specify later.

Setting $h=-1$ and $s_{\rm p} = {\rm sign}(p_z)$,  eq.\ \eqref{eq:CPodd_source} agrees with FH06 up to an overall sign.  A number of CP-odd source terms computed in FH06, proportional to $\mu_e$ and $\delta f_e$, were dropped during evaluation, since they are  higher order in gradients. 

%
\subsection{Moment expansion}
%

One could solve $\mu_{e,o}$ and $\delta f_{e,o}$ directly from Eqs.~\eqref{eq:CPeven} and~\eqref{eq:CPodd}. It is more economical however, to first reduce them to a set of moment equations. Because of their identical forms, the equations for both CP parities can be treated simultaneously. We introduce moments by integrating over $p$, weighted by $(p_z/E)^l$, and dividing by a normalization factor
\bea
	N_1 &\equiv& \int {\rm d}^3 p \,f'_{0w,\rm FD} = \gamma_w\int {\rm d}^3 p
\,f'_{0,\rm FD}\nn\\
	&\equiv& \gamma_w \hat N_1 = -\gamma_w\, \frac{2\pi^3}{3}\,T^2 \,,
	\label{N1eq}
\eea
where $f_{0,\rm FD}$ is the equilibrium distribution function for a massless fermion in the fluid frame.  It is convenient to normalize even the equations for a massive particle using this universal factor, so that when several species of particles are coupled through their interactions, the rates in the collision terms are related in a simple way between equations for different species. Then terms appearing in the fluid equations can be expressed as averages over phase space of the form
\be
  \langle X \rangle \equiv \frac{1}{N_1}\int {\rm d}^3p\, X \,.
\ee
In particular the integrals over $\delta f$ define the velocity perturbations
\be
 u_\ell \equiv  \Big\langle \! \left(\frac{p_z}{E}\right)^{\!\ell}\delta f \Big\rangle \,.
\label{dfeqs_2}
\ee
The $\ell$th moment of the evolution equation can then be written as
\be
\Big\langle \!\left(\frac{p_z}{E}\right)^{\!\ell} L \Big\rangle =  
\Big\langle \!\left(\frac{p_z}{E}\right)^{\!\ell}( {\cal S}  + \delta{\cal C}) \Big\rangle \,.
\label{eq:generic_moment}
\ee
Next we focus on the Liouville term~\eqref{eq:Liouville}, which contains important $v_w$ dependence.

%
\subsection{Liouville term}
%

Our goal is to reduce the system \eqref{eq:generic_moment} to a closed set of equations for $\mu$'s and the velocity perturbations~\eqref{dfeqs_2}. We  include only the two lowest moments, as has been done so far in the literature~\cite{Cline:2017qpe,Fromme:2006wx,Fromme:2006cm}.\footnote{Ref.\ \cite{Joyce:1994zn} also considered a temperature perturbation, but this is an ansatz for the distribution function, rather than a systematic expansion
in velocity moments.} Taking the zeroth and first moment of the Liouville operator we find
\bea
\langle L \rangle \! &=& \!\! -D_1 \mu' + u_1' 
            + v_w\gamma_w(m^2)^\prime Q_1 \mu 
\label{WKBeqs2a}
\\
\Big\langle \frac{p_z}{E}L\Big\rangle \!  &=& \!\! - D_2 \mu' + u_2' 
+v_w\gamma_w (m^2)^\prime  Q_2 \mu   \nn\\
&& + \,(m^2)^\prime\Big\langle \!\, \frac{1}{2E^2}\delta f \Big\rangle,
\label{WKBeqs2b}
\eea
where $^\prime$ again denotes $\partial_z$ except when acting on the distribution functions, where it denotes $\partial_{\gamma_w E}$, and we introduce the functions
\bea
D_\ell &\equiv& \Big\langle \!\left( \frac{p_z}{E}\right)^{\!\ell} \!f_{0w}^\prime \Big\rangle
\label{eq:D} \\
Q_\ell &\equiv& \Big\langle \!\left( \frac{p_z^{\ell-1}}{2E^\ell}\right) \!f_{0w}^{\prime\prime} \Big\rangle .
\label{eq:Q}
\eea
The $D$- and $Q$-functions are defined separately for bosons and for fermions, since the distribution function $f_{0w}$ differs in the two cases. In the small $v_w$-limit they reduce to the FH06-functions as  $D_1\rightarrow -v_wK_1^\HF$, $D_2\rightarrow K_4^\HF$ and $Q_1\rightarrow K_2^\HF$. The $Q_2$ term was however overlooked in FH06.

Eq.\ \eqref{WKBeqs2b} contains two problematic terms: $u_2$ is higher order in
the expansion than the order to which we are working, and the last term is not obviously related to velocity perturbations~\eqref{dfeqs_2}. To treat the first term we need to  introduce a {\em truncation} scheme, which relates $u_2$ to $u_1$. (More generally one should relate the $n$th moment to moments $u_\ell$ with $\ell<n$.) Here we  adopt a simple linear relation, henceforth  denoting $u_1 = u$:
%
$u_2 \equiv R u$ \,,
%
where $R$ is a function to be defined shortly. 

To define the last term in~\eqref{WKBeqs2b} we need a further {\em factorization} assumption. Following ref.\ ~\cite{Cline:2000nw} and FH06, for any ${\cal X}$ that does not correspond to a velocity perturbation we replace
\bea
\label{fact1}
 \langle {\cal X} \delta f\rangle &\rightarrow& [{\cal X} (E/p_z) ]u \\
    \ [{\cal X}]  &\equiv& \frac{1}{N_0} \int d^{\,3}p\, {\cal X} f_{0w}
\label{factorize}
\eea
where $N_{0}$ is another normalization factor,
\be
N_0 = \int d^{\,3}p\, f_{0w} = \gamma_w \int d^{\,3}p\, f_{0} \equiv  \gamma_w \hat N_0 \,.
\ee
Unlike $N_1$, $N_0$ is defined in terms of the massive distribution function $f_0$ of the particle under consideration.

In eq.\ (\ref{factorize}), it may happen that ${\cal X}$ does not have any power of $p_z$ to be canceled by the factor $E/p_z$ in eq.\ (\ref{fact1}).  Nevertheless the integral can be defined using the Cauchy principal value. In particular
\be
\left\langle \frac{1}{2E^2}\delta f \right\rangle \rightarrow \left[\frac{1}{2p_zE}\right]
u \equiv \bar R u.
\ee
After performing the singular angular integral using the principal value prescription, we find 
\be
\bar R = \frac{\pi}{\gamma_w^2 \hat N_0} \int_m^\infty {\rm d}E 
   \ln\left|\frac{p-v_w E}{p+ v_wE}\right| f_0.
\label{eq:barR}
\ee
This should reduce to $v_w \tilde K_6^\HF$ at leading order in $v_w$, but due to a mistake in the evaluation of $\tilde K_6^\HF$ in FH06 it does not. 

Following FH06 we use the factorization rule also to define the truncation scheme,
\be
u_2 = \Big\langle\! \left( \frac{p_z}{E} \right)^{\! 2}\! \delta f \Big\rangle \rightarrow \Big[\frac{p_z}{E}\Big] u \equiv R u.
\ee
where the bracket average $[\cdot]$ is defined in (\ref{factorize}).
Then $R$ becomes just the expectation value of the fluid velocity in the wall frame,
\be
R = -v_w,
\label{eq:R}
\ee
which is an exact result. Comparing with FH06 $R = v_w \tilde K_5^\HF$, this implies that $\tilde K_5^\HF = -1$ exactly. Although unstated in FH06, it is indeed the case. 

The factorization and the truncation rules are of course somewhat arbitrary. It is therefore reassuring that the $\bar R$-term has but a weak effect on solutions: toggling between the choice~\eqref{eq:barR} versus setting $\bar R \equiv 0$ changes the final baryon asymmetry at the level of a few per cent.  Moreover the definition~\eqref{eq:R} for $R$ is reasonable because $-v_w$ is roughly the ratio between adjacent source terms, order by order in the moment expansion.

%
\subsection{Sources and collision terms}
%

Assembling the previous results, and including the collision and source terms, the fluid equations can be presented in full detail. Defining a vector $w = (\mu,u)^T$, the general form of the two moment equations may be expressed as
\be
A w' + (m^2)'Bw = \! S + \delta C,
\label{WKBeqs}
\ee
where
\be
  A = \left( \!\! \begin{array}{cc}-D_1 &  1\\
	                               -D_2 &  R \end{array}\!\right)
\,,\quad 
  B = \left( \! \begin{array}{cc}v_w\gamma_w Q_1&   0 \\
	                               v_w\gamma_w Q_2 &  \bar R \end{array}\!\right)
\label{eq:AB}
\ee
and $S = (S_1,S_2)^T$ with $S_1 = \langle{\cal S}\rangle$ and $S_2 = \langle(p_z/E){\cal S}\rangle$ and similarly for the $\delta C$ vector. The
form (\ref{eq:AB}) is generic to both CP-even and CP-odd sectors, which are only distinguished by their respective source terms.

Let us consider the source terms first. In the CP-even sector one finds using~\eqref{eq:CPeven_source}
\be
S^e_\ell = v_w \gamma_w (m^2)^\prime Q^e_\ell,
\ee
with the definition
\be
 Q^e_\ell \equiv \Big\langle \frac{p_z^{\ell-1}}{2E^\ell} f_{0w}^\prime \Big\rangle .
\ee 
In the small-$v_w$ limit one finds $Q_2^{e} \rightarrow K_3^\HF$. 

In the CP-odd sector,  using eq.\ \eqref{eq:CPodd_source} similarly gives
\be
S^o_{h\ell} = -v_w \gamma_w h \big[ (m^2\theta^\prime)^\prime Q^{8o}_\ell - (m^2)^\prime m^2\theta^\prime 
Q^{9o}_\ell \big] ,
\label{sterms}
\ee
where the coefficient functions are
\bea
 Q^{8o}_\ell &\equiv& \Big\langle \frac{s_{\rm p}p_z^{\ell-1}}{2E^\ell E_z} f_{0w}^\prime 
\label{eq:Q8} 
\Big\rangle  
\\
 Q^{9o}_\ell &\equiv& \Big\langle \frac{s_{\rm p}p_z^{\ell-1}}{4E^{\ell+1}E_z} 
 \Big( \frac{1}{E}f_{0w}^\prime -\gamma_w f_{0w}^{\prime\prime} \Big) \!\Big\rangle ,
\label{eq:Q9} 
\eea 
with $s_{\rm p}$ defined in~\eqref{eq:shel}. Setting $s_{\rm p} \rightarrow {\rm sign}(p_z)$ one finds that $Q^{8o}_2 \rightarrow K_8^\HF$ and $Q^{9o}_2 \rightarrow K_9^\HF$ in the small $v_w$-limit. Moreover, in previous work, the approximation $S_1 = 0$ was always made, because it is $O(v_w^2)$. For large velocities there is no hierarchy between $S_1$ and $S_2$ and one must include both sources.

It remains to consider the collision integrals. Both phase space averages $\delta C_1 \equiv \langle \delta {\cal C}\rangle$ and $\delta C_2 \equiv \langle (p_z/E)\delta {\cal C}\rangle$ are normalized using $N_1^{-1}$, eq.\ (\ref{N1eq}). The collision term moments are derived following appendix A of ref.\ \cite{Cline:2000nw},
\bea
\delta  C_1 &=& K_0 \sum_i\Gamma_i \sum_j s_{ij}\frac{\mu_j}{T}\,,\nn\\
\delta  C_2 &=& - \Gamma_{\rm tot}\,u - v_w \delta C_1\,.
\label{eq:collision_terms}
\eea
Here $s_{ij} = 1$ ($-1$) if the corresponding species is in the initial (final) state in the interaction with rate $\Gamma_i$, and $\Gamma_{\rm tot} = \sum_i\Gamma_i$ is the total interaction rate, including elastic channels that do not contribute to the sum in $ \langle \delta {\cal C}\rangle$. The normalization factor
\be
K_0 \equiv - \langle f_{0w} \rangle = -\frac{\hat N_0}{\hat N_1}
\ee
was neglected in FH06. For a massless fermion, for example, $K_0\cong 1.1$, Eqs.~\eqref{eq:collision_terms} are valid for both the CP-even and the CP-odd cases. 

Eqs.~\eqref{WKBeqs} obviously depend on a large number of coefficient functions: $D_{\ell}$, $Q_{\ell}$, $R$, $\bar R$, $Q^{e}_{\ell}$, $Q^{8o}_{\ell}$, $Q^{9o}_{\ell}$ and $K_0$. Most of these  depend on the wall velocity $v_w$ and the dimensionless ratio $x = m/T$. However they are {\it universal} and model independent. In practice, we compute them on a grid of $x$ and $v_w$ values and spline fit them. Explicit expressions are given for all the integrals definng
them in appendix~\ref{sec:app_explicit_forms}.

\begin{figure}[t]
\hspace{-0.4cm}
\centerline{\includegraphics[width=0.95\hsize]{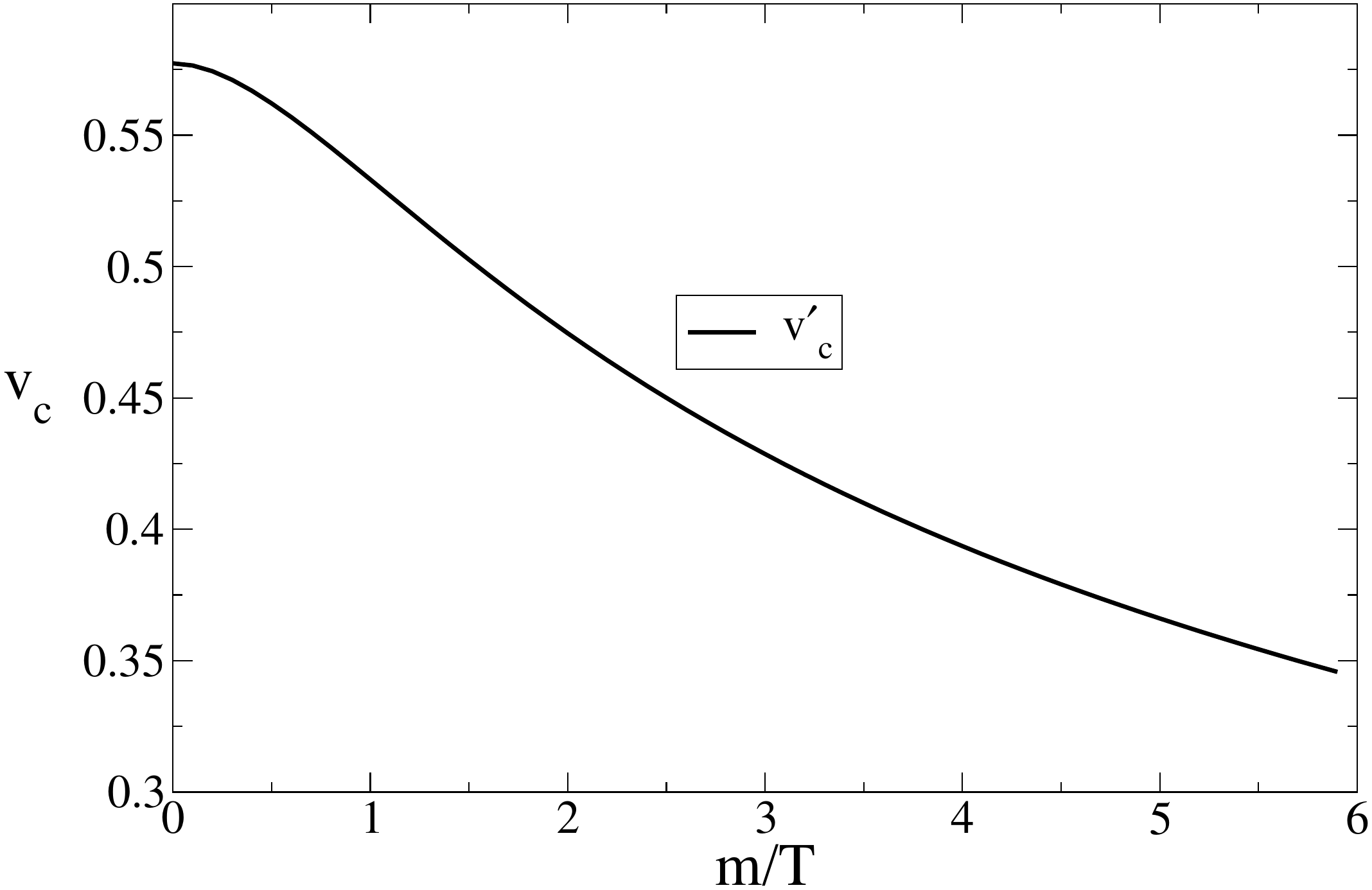}}
\caption{Naive prediction  for the critical
wall velocity from FH06 equations, as function of $m/T$.
The correct value, using the full $v_w$-dependence of the $D_\ell$ functions, is $v_c = 1$.}
\label{HFvs}
\end{figure}

%
\subsection{Critical speed predictions}
\label{sec:critical_speed}
%

There is a widespread notion, apparently originating from ref.\ \cite{Huet:1995sh}, that diffusion is inefficient for wall speeds exceeding the plasma sound speed. This would mean in particular that EWBG would not be feasible for detonation walls, corresponding to very strong phase transitions, often invoked in the context of gravitational wave production. This assertion is not true, as we shall show, but it turns out that the FH06 equations are, quite fortuitously, consistent with the false assumption.

We noted that the fluid equations can be written in the matrix form $A w' = F[w]$, where  $A$ is given in Eq.\ \eqref{eq:AB}, while in FH06 the $A$ matrix is
\bea
  A_\HF = \left(\begin{array}{cc} v_w K_1^\HF &  1\\
	 -K_4^\HF &  -v_w \end{array}\right),
\label{FHeqs}
\eea
setting $\tilde K_5^\HF=-1$ as mentioned above.
One can solve for the value of $v_w$ where $A$ becomes singular (noninvertible) using ${\rm det}(A)=0$. If a solution exists for $v_w < 1$ it implies a critical speed $v_c$ beyond which diffusion is quenched. The exact prediction using the $A$-matrix in~\eqref{eq:AB} gives (recalling that $R = -v_w$) 
\be
  v_c = -\left( {D_2 \over D_1}\right)_{v_w=v_c} \quad \Rightarrow \quad v_c = 1\,,
\label{eq:vc_exa}
\ee
whereas the approximate FH06-condition gives a different velocity
\be
v_c^\prime = \left|{K_4^\HF \over K_1^\HF}\right|^{1/2}_{v_w=0}\!.
\label{eq:vc_app}
\ee

The dependence of $v_c^\prime$ on $m/T$ as obtained in the FH06-case~\eqref{eq:vc_app} is shown in fig.\ \ref{HFvs} for a Fermi-Dirac distribution. (The corresponding curve for bosons look similar.) For light particles the quench limit is maximal and very close to the sound speed, but this is a mere coincidence due inappropriate use of the small $v_w$-approximation. Indeed, from~\eqref{eq:vc_exa}, employing full $v_w$-dependent function, we find that $v_c=1$, in accordance with the arguments given in section~\ref{sect:vw}. Thus diffusion efficiency should go to zero smoothly as $v_w \rightarrow 1$, with no particular features at the sound speed, $v_w = v_s$. We will show that this indeed is the case.

%
\section{Phenomenological model}
\label{sec:model}
%

To illustrate the consequences of our improved transport equations, we will compute the baryon
asymmetry that they predict in a prototypical model that gives rise to EWBG, where the top quark mass has a $z$-dependent CP-violating phase in the bubble wall. The mass term can be written as
\be
	m_t(z) \left(\bar t_\L e^{i\theta(z)} t_\R + \bar t_\R e^{-i\theta(z)} t_\L \right)
\ee
in terms of the chiral components, where $m_t = y_t v(z)$ and $v(z)$ is the Higgs VEV that varies spatially within the wall.  It can occur in two-Higgs-doublet models, or in singlet plus doublet models where a dimension-5 operator
like $i(s/\Lambda) \bar Q_3 H t_\R$ contributes a phase to the top mass, if $s$ also gets a VEV in the bubble wall.  In such a model, the effective top quark mass term takes the form
\be
	y_t h(z)\,\bar t_\L \left( 1 + i {s(z)\over \Lambda}\right) t_\R + {\rm H.c.}
\label{mteq}
\ee
which implies
\bea
	m_t(z) &=& y_t h(z)\sqrt{1 + s^2(z)/\Lambda^2}\nn\\
	\theta(z) &=& \tan^{-1}{s(z)\over\Lambda}\,.
\eea

Here we will not consider the CP-even equations, which would be relevant for computing the wall speed and shape. Instead, we concentrate on the CP-odd sector and take a phenomenological approach, where $v_w$ is treated as a free parameter, and the VEVs $h(z)$, $s(z)$ are modeled as
\bea
	h(z) &=& {v_n\over 2}\left(1 - \tanh {z\over L_w}\right)\nn\\
	s(z) &=& {w_n\over 2}\left(1 + \tanh {(z-\delta_w)\over L_s}\right).
\label{wall-profile}
\eea
We are primarily interested in the $v_w$-dependence of the results and 
therefore choose fiducial values for the other parameters,
\bea
&& v_n = \sfrac{1}{2} w_n = T_n,\quad  \Lambda = 1\,{\rm TeV}\nn\\
&& L_w = L_s = {5\over T_n},\quad \delta_w = 0 \,,
\label{eq:fiducial_model}
\eea
in terms of the nucleation temperature, taken to be $T_n = 100$ GeV. 

%
\subsection{Fluid equation network}
%

With the tools and notation developed in the previous section one can express the complicated equation network in a compact form.
Our system consists of four particle species\footnote{We differ from the notation of FH06 by keeping track of the asymmetry in the right helicity sector with $t_R$ instead of its
conjugate $t_\R^c$. The CP-odd asymmetries of the two species have a relative sign, $\mu_{t_R} = - \mu_{t_\R^c}$, since each CP-odd variable represents the difference between the two C-conjugate species. Likewise, the source for $\mu_{t_R}$ corresponds to $h=1$, which is opposite to that for $\mu_{t_\R^c}$.} strongly coupled by the top-Yukawa interactions: left and right helicity tops are respectively denoted by $\tL$ and $\tR$, left helicity bottom  by $b_\L$ and the complex Higgs particle by $h$. We neglect the small difference between helicity and chirality of the fermions here. There are eight dependent variables, combined into four 2-vectors $w_i = (\mu_{oi},u_{oi})^T$ for $i=\tL,\bL,\tR,h$, which obey
\def\phm{{\phantom{-}}}
\bea
A_{t} w_{\tL}' + m_t^{2\,\prime}B_t\, w_{\tL} - \delta C_{\tL}  \!&=&\!\! \phm  S_t, \nn \\
A_{b} w_{\bL}' + m_b^{2\,\prime}B_b w_{b_L}   - \delta C_{\bL}  \!&=&\!\! \phm  S_b, \nn \\
A_{t} w_{\tR}' + m_t^{2\,\prime}B_t w_{\tR}   - \delta C_{\tR}  \!&=&\!\!  -    S_t, \nn \\
A_{h} w_{h\;}' + m_h^{2\,\prime}B_h w_{h\;}   - \delta C_{h\;}  \!&=&\!\! \phm  0,
\label{eq:network}
\eea
where the $A-$ and $B$-matrices are defined in~\eqref{eq:AB} and the sources are $S_i = (S_{i1},S_{i2})^T$,  $S_{i\ell}$  given by equation~\eqref{sterms} taking $h\equiv -1$ for the left-handed fermions. In practice $S_b$ is neglible due to the smallness of the bottom Yukawa coupling.

In addition to top-Yukawa interactions, we account for the $W$ boson interactions that tend to equalize the $t_\L$ and $b_\L$ chemical potentials, the strong sphalerons, top mass insertions (helicity flips) that damp the combination $\mu_{t_\L} \!-\! \mu_{t_\R}$ $(\mu_h)$, and Higgs damping from electroweak symmetry breaking. These are the same collision terms as in FH06. Explicitly $\delta {\cal C}_i \equiv (K^i_0\, \delta\overline{\cal C}^i_1,\delta {\cal C}^i_2)^T$, where
\bea
\label{Cs}
\delta \overline{\cal C}^{\tL}_1 \!&=&  \Gamma_{y}\, (\mu_\tL \!-\! \mu_\tR \!+\! \mu_h) 
                     + \Gamma_{\rm m}\, (\mu_\tL \!-\! \mu_\tR)  \nn\\
                 &+& \Gamma_{\rm W} (\mu_\tL\!-\!\mu_\bL) + \tGamma_{\rm SS}[\mu_i]
\nn \\
\delta \overline{\cal C}^{\bL}_1 \!&=&  \Gamma_{y}\, (\mu_\bL \!-\! \mu_\tR \!+\! \mu_h) \nn\\
                 &+& \Gamma_{\rm W} (\mu_\bL\!-\!\mu_\tL) + \tGamma_{\rm SS}[\mu_i]
\nn\\
\delta \overline{\cal C}^{\tR}_1 \!&=& -\Gamma_{y}\, (\mu_\tL \!+\! \mu_\bL \!-\! 2\mu_\tR \!+\! 2\mu_h) \nn \\
&+&  \Gamma_{\rm m}\, (\mu_\tR \!-\! \mu_\tL) \!-\! \tGamma_{\rm SS}[\mu_i]
\nn\\
\delta \overline{\cal C}^h_1 \!&=& \tGamma_{y}\, (\mu_\tL \!+\! \mu_\bL \!-\! 2\mu_\tR \!+\! 2\mu_h) + \Gamma_h \mu_h \,,
\label{eq:collision_terms2}
\eea
and $\delta {\cal C}^i_2 = -\Gamma^i_{\rm tot} u_i - v_w K^i_0\delta \overline{\cal C}^i_1$. Explicit equations for 
light quarks are not needed, even though their chemical potentials appear in the strong sphaleron rate $\tGamma_{\rm SS}[\mu_i] = \Gamma_{\rm SS}\sum_q(\mu_{q_\L} \!-\!\mu_{q_\R})$, since their chemical potentials can be determined analytically. Light quarks are activated only by strong sphalerons, and in the approximation of no Yukawa mixing, $\mu_{q_\R}=-\mu_{q_\L}$ for all light species. Then using baryon number conservation (neglecting electroweak sphalerons, which are slow on the relevant time scale), $B = \sum_q (n_q-\bar n_q) = 0$, one finds
\be
\mu_{q_\L} = - \mu_{q_\R} = D_0^t\mu_{t_\L} + D_0^b\mu_{b_\L} + D_0^t\mu_{t_\R},
\label{eq:lightquarks}
\ee
where $D_0 = \langle f_{0w}^\prime \rangle$ is a special case of the function~\eqref{eq:D} with $\ell = 0$, identical to  $K_1^\HF$ for all $v_w$, as can be shown by partial integration. Using~\eqref{eq:lightquarks} the strong sphaleron rate can be written as
\bea
\tGamma_{\rm SS}[\mu_i] 
= \Gamma_{\rm SS} \big( (1+9D_0^t) \mu_\tL  \!\!\!&+&\!\! (1+9D_0^b)\mu_\bL \nn\\
 &-& \!\!(1-9D_0^t)\mu_\tR \big) . 
\label{eq:strongsph}
\eea

Inelastic collisions induce mixing between the particle species. Eqs.~\eqref{eq:network} are nevertheless linear in $\mu_i$ and $u_i$, and can be written in the compact matrix form
\be
  A U' - \Gamma U = S
\label{deqs}
\ee
where $U^T \equiv (w_\tL^T,w_\bL^T,w_\tR^T,w_h^T)$, $A = {\rm diag}(A_\tL, A_\tR, A_\bL,$ $A_h)$ is tridiagonal, and the matrix $\Gamma$ combines the $m_i^{2\,\prime}B_i$ and collision terms. The source vector is $S^T = (S_t^T,S_b^T,-S_t^T,S_h^T)$ with $S_i$ defined in Eq.~\eqref{eq:network}. Because of its block structure, $A$ is easily inverted to yield
\be
  U' = A^{-1}\Gamma U + A^{-1}S \,.
\ee
This system is best solved using relaxation methods~\cite{cash} since shooting tends to be unstable. The $8\!\times\! 8$-matrix $A^{-1}\Gamma$ is the Jacobian of the differential equation network and its eigenvalues' signs distinguish the growing and decaying modes at the boundaries. This information may help to improve the numerical stability in more complicated systems.

Once the chemical potentials for the perturbations are determined, the baryon asymmetry follows from integrating them in the sphaleron rate equation. Following for example ref.~\cite{Cline:2011mm} (but including the full Lorentz-covariant relations) we find:\footnote{To our knowledge, the $1/\gamma_w$ factor in front has been omitted in previous literature.  It arises from the change of variable
$dt\to dz/(v_w\gamma_w)$ with $z$ in the rest frame of the wall.}
\be
	\eta_B = {405\,\Gamma_{\rm sph}\over 4\pi^2 v_w\gamma_w g_* T}\int dz\, 
	\mu_{\!B_{\rm L}}f_{\rm sph}\,e^{-45\Gamma_{\rm sph}|z|/4v_w\!\gamma_w}.
\label{etab}
\ee
The seed asymmetry in eq.\ (\ref{etab}) is the chemical potential for left-handed baryon number, $\mu_{B_{\rm L}} = \sfrac{1}{2}\sum_q \mu_{q_{\rm L}}$, which can be written in terms of $\mu_\tL, \mu_\bL$ and $\mu_\tR$ using baryon number conservation:
\be
\mu_{\!B_\L} = \sfrac12(1 + 4 D_0^t)\mu_\tL + \sfrac12(1 + 4 D_0^b)\mu_\bL + 2D_0^t \mu_{\tR}.
\ee
The function $f_{\rm sph}(z) = {\rm min}(1,2.4\frac{\Gamma_{\rm sph}}{T}e^{-40h(z)/T})$ is designed to smoothly interpolate between the sphaleron rates in the broken and unbroken phases. $g_*$ is the number of degrees of freedom in the heat bath; in the standard model  $g_*= 106.75$.

%
\section{Comparison to FH06}
\label{sec:comparisonFH06}
%

We can now compare our improved fluid equations~(\ref{eq:network}-\ref{eq:collision_terms2}) to those of FH06. The only difference between the two lies in the definition of the various coefficient functions, which we have renamed at the same time correcting and generalizing them to arbitrary wall velocities. To facilitate the comparison the results are collected in a dictionary translating between the two naming schemes in table~\ref{tab:dictionary}. Our equations agree with those of FH06,  when one assumes $h \rightarrow -1$ and 
$s_{\rm p}\rightarrow {\rm sign}(p_z)$ in the sources and replaces the coefficent functions according to table~\ref{tab:dictionary}.
 
\def\phmm{{\phantom{m}}}
\begin{table}[ht]
\begin{center}
\begin{tabular}{lcr}
CK                    & \phmm\phantom{*} &                   FH06 \\
\toprule
$D_0(x)          $    & \phmm          = &   $       K_1^\HF (x)$ \\
$D_1(x,v_w)      $    & \phmm          = &   $  -v_w K_1^\HF (x)$ \\
$D_2(x,v_w)      $    & \phmm\phantom{*} &   $       K_4^\HF (x)$ \\
$Q_1(x,v_w)      $    & \phmm\phantom{*} &   $       K_2^\HF (x)$ \\
$Q_2(x,v_w)      $    & \phmm         !  &   $                 0$ \\
$R = -v_w        $    & \phmm         =  &   $ v_w\tilde K_5^\HF$ \\
$\bar R(x,v_w)   $    & \phmm         !! &   $\tilde K_6^\HF (x)$ \\
$Q^e_1(x,v_w)    $    & \phmm          ! &   $                 0$ \\
$Q^e_2(x,v_w)    $    & \phmm\phantom{*} &   $       K_3^\HF (x)$ \\
$Q^{8o}_1(x,v_w) $    & \phmm          ! &   $                 0$ \\
$Q^{8o}_2(x,v_w) $    & \phmm\phantom{*} &   $       K_8^\HF (x)$ \\
$Q^{9o}_1(x,v_w) $    & \phmm          ! &   $                 0$ \\
$Q^{9o}_2(x,v_w) $    & \phmm\phantom{*} &   $       K_9^\HF (x)$ \\
$K_0(x)          $    & \phmm         !! &   $                 1$ \\
\toprule
\end{tabular}
\end{center}
\vskip -0.3 truecm
\caption{A dictionary between the CK (this work) and the FH06 functions, depending upon $x=m/T$ and wall velocity $v_w$. They generally differ from each other at large $v_w$. Functions that are equivalent are marked by an equality sign in the middle column. The double exclamation mark indicates functions that do not agree even for small $v_w$ and single exclamation marks signal the source terms omitted in FH06.}
\label{tab:dictionary}
\end{table}%

For the interaction rates we use the values given in~\cite{Fromme:2006cm}:
$\Gamma_{\rm sph}=1.0\times 10^{-6}T$, $\Gamma_{\rm SS}=4.9\times 10^{-4}T$, 
$\Gamma_y=4.2\times 10^{-3}T$, $\Gamma_m=m^2_t/(63T)$ and $\Gamma_h=m^2_W/(50T)$, where top mass is as given in~\eqref{mteq} and $m^2_W \equiv g^2h(z)^2/4$. Furthermore the total interaction rates were defined as $\Gamma^i_{\rm tot} = K_{4,i}^\HF/(D_i K_{1,i}^\HF)$ with a quark diffusion constant $D_q=6/T$ and a Higgs diffusion constant $D_h = 20/T$. The numerical impact of the Higgs and bottom masses is found to be quite small, and following FH06 we take them to be massless.  Many of these rates have been quite roughly estimated, in some cases going back to the early reference \cite{Huet:1995sh}, and deserve to be updated.  We hope to make
better determinations in an upcoming paper.

\begin{figure}[t]
\hspace{-0.4cm}
\includegraphics[width=0.95\hsize]{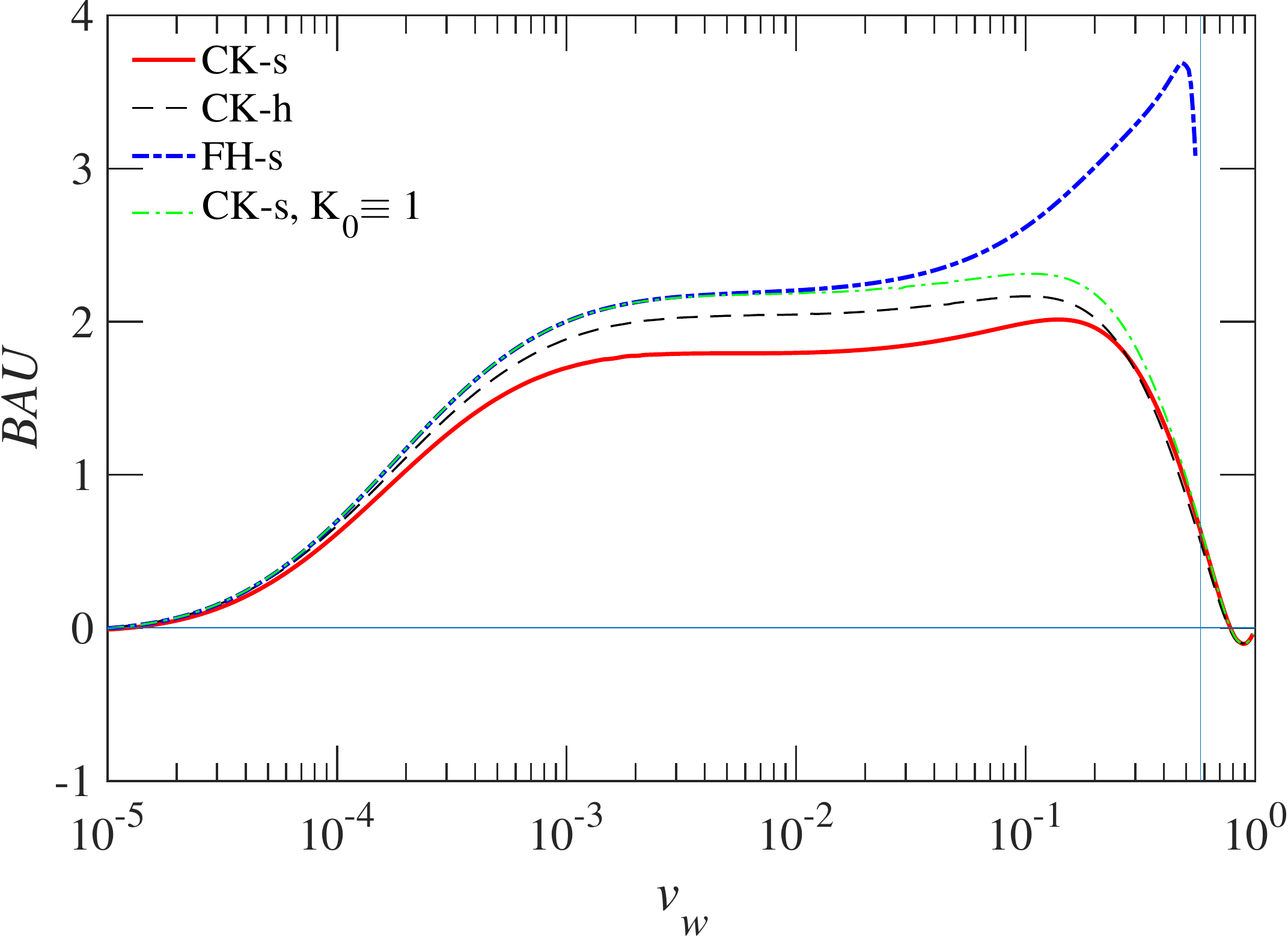}\\ \vskip0.5truecm \hskip-0.3truecm
\includegraphics[width=0.95\hsize]{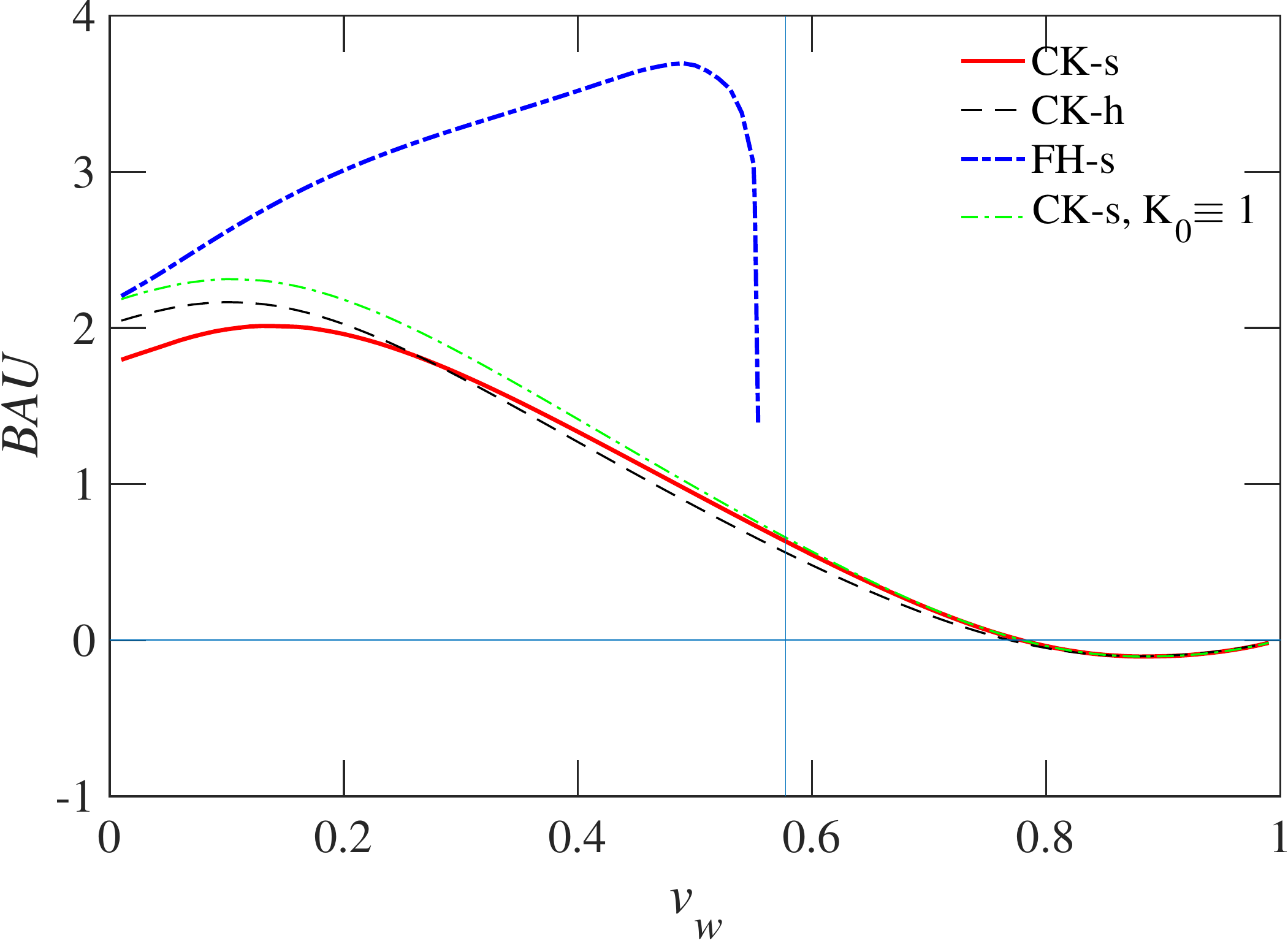} 
\caption{Predicted baryon asymmetry in units of observed asymmetry for the fiducial profile as a function of the wall velocity $v_w$. From the logarighmic scale plot (upper panel) one can appreciate the good agreement at small $v_w$. Note the vanishing of BAU for $v_w \lsim 10^{-5}$ due to the onset of thermal equilibrium. The linear scale (lower panel) expands the large $v_w$ region more relevant for strong transitions. Thin vertical line shows the sound speed $v_s=1/\sqrt{3}$.}
\label{fig:F0H6_comparison}
\end{figure}

We display dependences of the predicted baryon asymmetry of the universe normalized to the observed value, BAU $\equiv \eta_B/\eta_{B,\rm obs}$, in fig.~\ref{fig:F0H6_comparison}.  In both panels the thick red solid lines labeled ``CK-$s$'' correspond to the improved fluid equations with the spin-$s$ source, where we set $s_{\rm p}\rightarrow {\rm sign}(p_z)$ and $h=-1$ in  eqs.~(\ref{sterms}-\ref{eq:Q9}). The thick dash-dotted blue lines labeled ``FH-$s$'' correspond to the same spin-$s$ source, but using the FH06 equations\footnote{We switched for the sign of the source in FH06 however, so that the sign of the BAU matches in both cases.}. Thin dashed green lines labeled ``CK-$s$, $K_0=0$'' correspond to the case where we set the $K_0$-function to unity in the otherwise accurate equations. The thin black dashed lines labeled ``CK-$h$'' correspond to the improved fluid equations with the helicity source, still taking $h=-1$, but with $s_{\rm p}$ given by eq.~\eqref{eq:shel}.

Clearly all approximations agree very well for $v_w\ll 1$ as  expected, since the two sets of functions largely agree in the small $v_w$-limit; for $v_w\lsim 0.01$, the only significant difference between the CK and the FH06 solutions comes from $K_0$. For larger $v_w$ the predictions  differ significantly,  in accord with our general arguments. In particular, the FH prediction plummets as $v_w$ approaches the sound speed $v_s = 1/\sqrt{3}$, shown by the thin vertical line in the plots. The more exact treatment on the other hand does nothing special near  $v_w = v_s$; as expected the BAU smoothly decays as $v_w\to 1$.

Using the spin-$s$ source corresponds to identifying chirality with the eigenstates of spin in the $z$-direction, in the frame where the parallel momentum of the state vanishes, whereas the helicity source identifies chirality with helicity. The difference between the two is found to be small, due to the two bases becoming degenerate in the massless limit; in our example all fermions are massless in front of the wall.

\begin{figure}[t]
\hspace{-0.4cm}
\includegraphics[width=0.95\hsize]{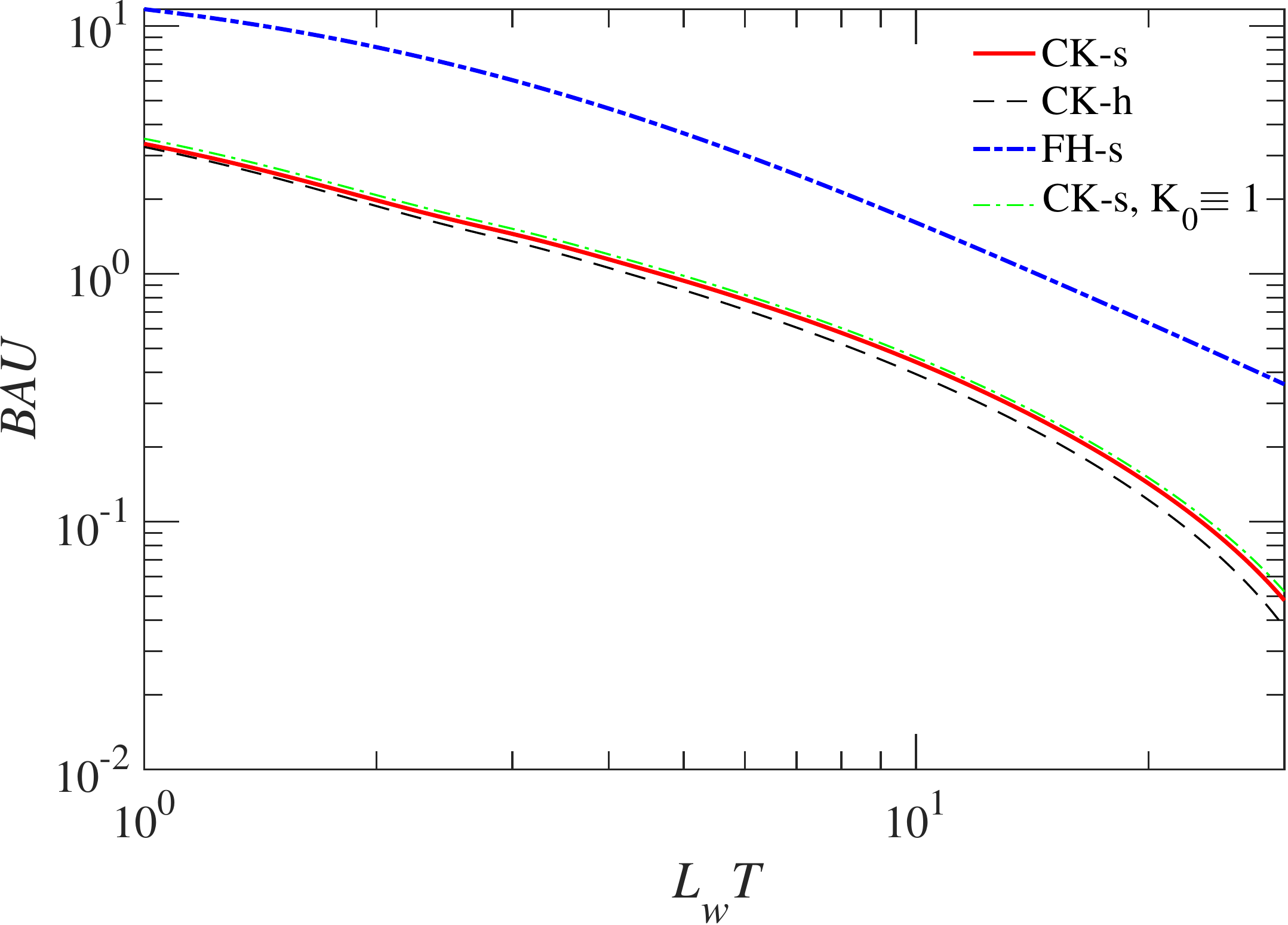}
\caption{Predicted baryon asymmetry(in units of observed asymmetry for the fiducial profile. as a function of the Higgs wall width $L_w$ for fixed $v_w = 0.5$.}
\label{fig:F0H6_comparison_2}
\end{figure}

In figure~\ref{fig:F0H6_comparison_2} we show the dependence on $L_w$ with $v_w =0.5$ held fixed. The FH prediction is substantially higher than the accurate value and its ratio to the correct solution remains nearly constant. To summarize, our results and those of FH agree reasonably well for small $v_w$, but the improved fluid equations should be used for $v_w \gsim 0.1$ to get accurate results, and for $v_w \gsim v_s$ they are essential, since the FH equations incorrectly predict a vanishing BAU. 

%
\section{Comparison to other formalisms}
\label{sec:fcomp}
%

There has been a long-standing divide among practitioners of electroweak baryogenesis as to which transport equations to use; yet no systematic comparison between them has been made in the literature.  We undertake to do so in this section, continuing with the ansatz for the wall profiles and spatially varying top quark mass (\ref{mteq},\ref{wall-profile}) introduced previously.  

\begin{figure}[t]
\hspace{-0.4cm}
\includegraphics[width=0.95\hsize]{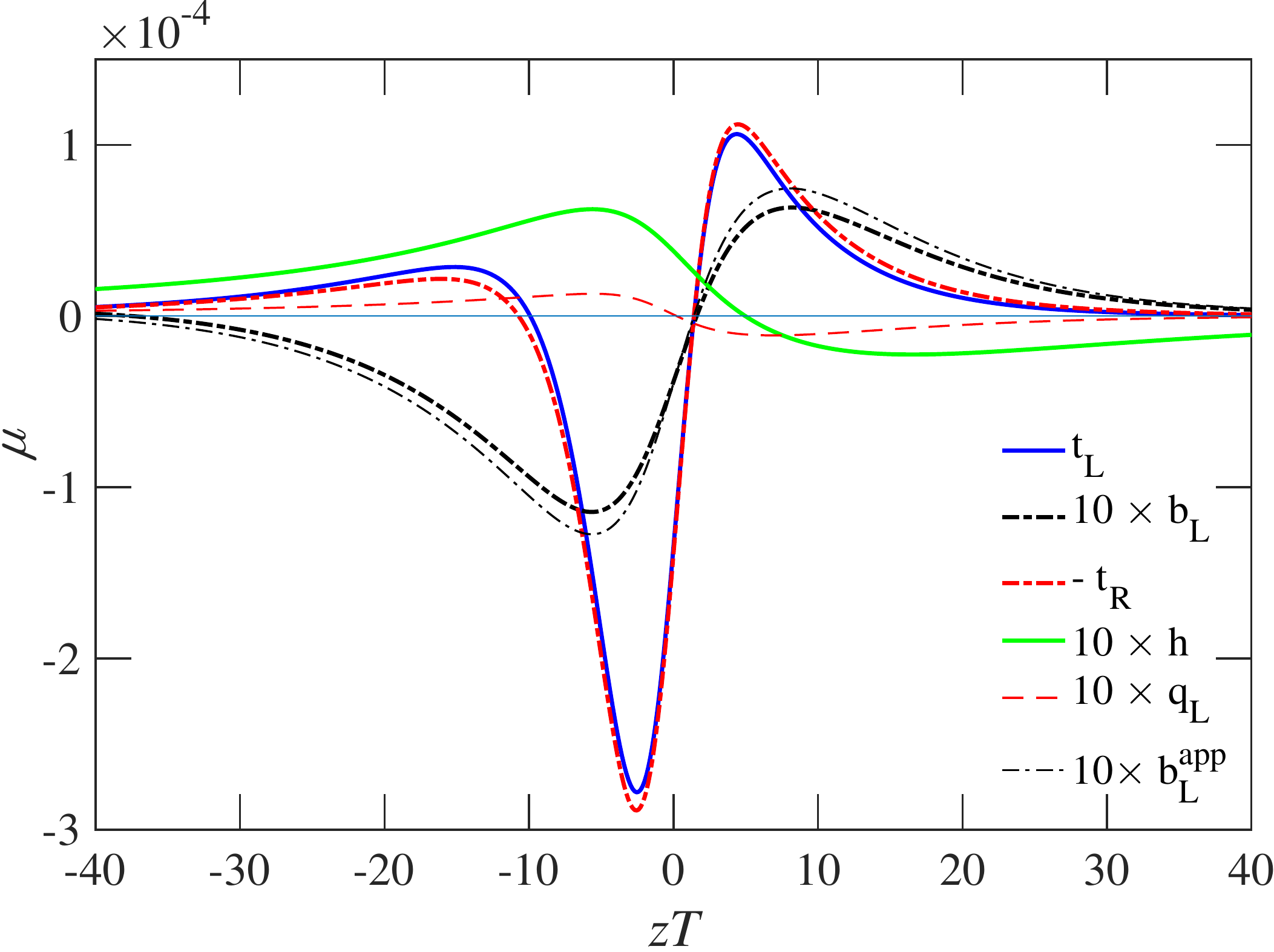}
\caption{Chemical potentials from the improved fluid equations for the fiducal case and for the spin-$s$ source. In addition to the chemical potentials in the network, we show light quark chemical potential $\mu_q$ corresponding to eq.~\eqref{eq:lightquarks} and the approximation $\mu_{b_{\rm L}^{\rm app}} \equiv -(D_0^t/D_0^b)(\mu_\tL+\mu_\tR)$.}
\label{fig:profiles}
\end{figure}

The VEV insertion formalism is derived at the level of the integrated particle densities, which is equivalent to the formalism introduced in refs.~\cite{Cohen:1994ss,Huet:1995sh}, consisting of coupled second-order diffusion equations for the local particle densities, in matrix form,
\be
  D\mu'' + v_w\mu' - \delta C[\mu] = S.
\label{HNeq}
\ee
Here $\mu = (\mu_{t_\L},\mu_{b_\L},\mu_{t_\R},\mu_h)^T$ and $S = (S_t,0,-S_t,0)^T$, $D={\rm diag}(D_q,D_q,D_q,D_h)$ is a diagonal matrix of diffusion coefficients and $\delta C[\mu]$ is the inelastic collision integrals. The unsourced equation for $\mu_{b_\L}$ is usually omitted in the literature~\cite{Chung:2009cb,Tulin:2011wi}. In this case baryon number conservation (as discussed above) leads to conditions $\mu_{q_\L} = \mu_{q_\R}$ and $\mu_{b_\L} = -(D_0^t/D_0^b)(\mu_{t_\L}+\mu_{t_\R})$~\cite{Huet:1995sh,Chung:2009cb}. In fact this is a reasonable approximation, as one can see from figure~\ref{fig:profiles}, where we plot the chemical potentials for our fiducial case~\eqref{eq:fiducial_model} and for the spin-$h$ source using our improved fluid equations. For the purpose of comparing the formalisms however, we have included the $b_L$ degrees of freedom in the diffusion equation networks. 

In order to compare the system~\eqref{HNeq} to our improved equations~\eqref{eq:network}, we need to find the equivalent source term to use in~\eqref{HNeq}. The standard way to do this~\cite{Cline:2000nw} is by eliminating $u$ from the WKB equations~\eqref{WKBeqs}, neglecting all $(m^2)'\mu$ and $(m^2)'u$ terms in the derivation. It is easy to show that this procedure yields the following results adequate for the WKB-picture\footnote{One should not confuse the diffusion coefficients with the fluid equation coefficient functions $D_i$. The latter are distinguished from the former by the fact that they are always associated with a numeral index.}\,\footnote{Note that the numerator in the diffusion coefficient $D_{\rm WKB}$ is just the determinant of the matrix $A$ in fluid equations (\ref{WKBeqs}). Thus, the critical speed condition discussed in section~\ref{sec:critical_speed} corresponds to zero diffusion length. For $v_w > v_c$ the diffusion length would be negative, which is of course unphysical. This is why the FH06 solutions go to zero as $v_w\rightarrow v_c$. However, in the improved equations $D_{\rm WKB}$ and hence the diffusion length remains positive until $v_w = 1$.}

\bea
    D_{\rm WKB} &=& {D_2 - v_w^2 D_0\over D_0\Gamma_{\rm tot}},\nn\\
	S_{\rm WKB} &=& {S_1\over D_0} - {v_w S_1' + S_2'\over D_0\Gamma_{\rm tot}},\nn\\
    \delta C_{\rm WKB} &=& {K_0\over D_0}\delta \bar{\cal C}[\mu]\,.
\label{source_conv}
\eea
where $\delta \bar{\cal C}[\mu]$ terms are given in~\eqref{eq:collision_terms2}.
In the small $v_w$-limit $D_{\rm WKB}^\alpha \rightarrow D_\alpha$ and $S^\alpha_{\rm WKB} \rightarrow D_\alpha S_{\alpha,2}^\prime/\langle v_z^2\rangle$  for each species $\alpha = q,h$, in agreement with ref.~\cite{Cline:2000nw}.

The VEV-insertion formalism predicts a very different form for the source term in (\ref{HNeq}). We give a detailed derivation in appendix \ref{appB}.  The result, normalized as in eq.~(\ref{HNeq}), is
\be
  S^t_{\rm VEV} = v_w{N_c I\over 2\pi^2 D_0\,T}\, m_t^2(z)\theta'(z)
\label{Steq}
\ee
where $N_c=3$ is the number of colors of the top quark and $I \cong 0.4$ is an integral given in appendix \ref{appB}. 

As for the diffusion term, no dependence of $D$ on $v_w$ is considered in the earlier literature in the VEV insertion approach. Accordingly, we will use
\be
D^\alpha_{\rm VEV} = D_\alpha
\label{eq:D_vev}
\ee
As mentioned above, we employ the same equation network for the WKB- and the VEV insertion mechanisms, so that slightly upgrading the network of ref.~\cite{Tulin:2011wi} to include
$b_L$, we set 
\be 
\delta C_{\rm VEV} = \delta \bar {\cal C}[\mu] \,.
\label{eq:C_vev}
\ee
where as before $\delta \bar{\cal C}[\mu]$ correspons to eq.~\eqref{eq:collision_terms2}.

\begin{figure}[t]
\hspace{-0.4cm}
\includegraphics[width=0.95\hsize]{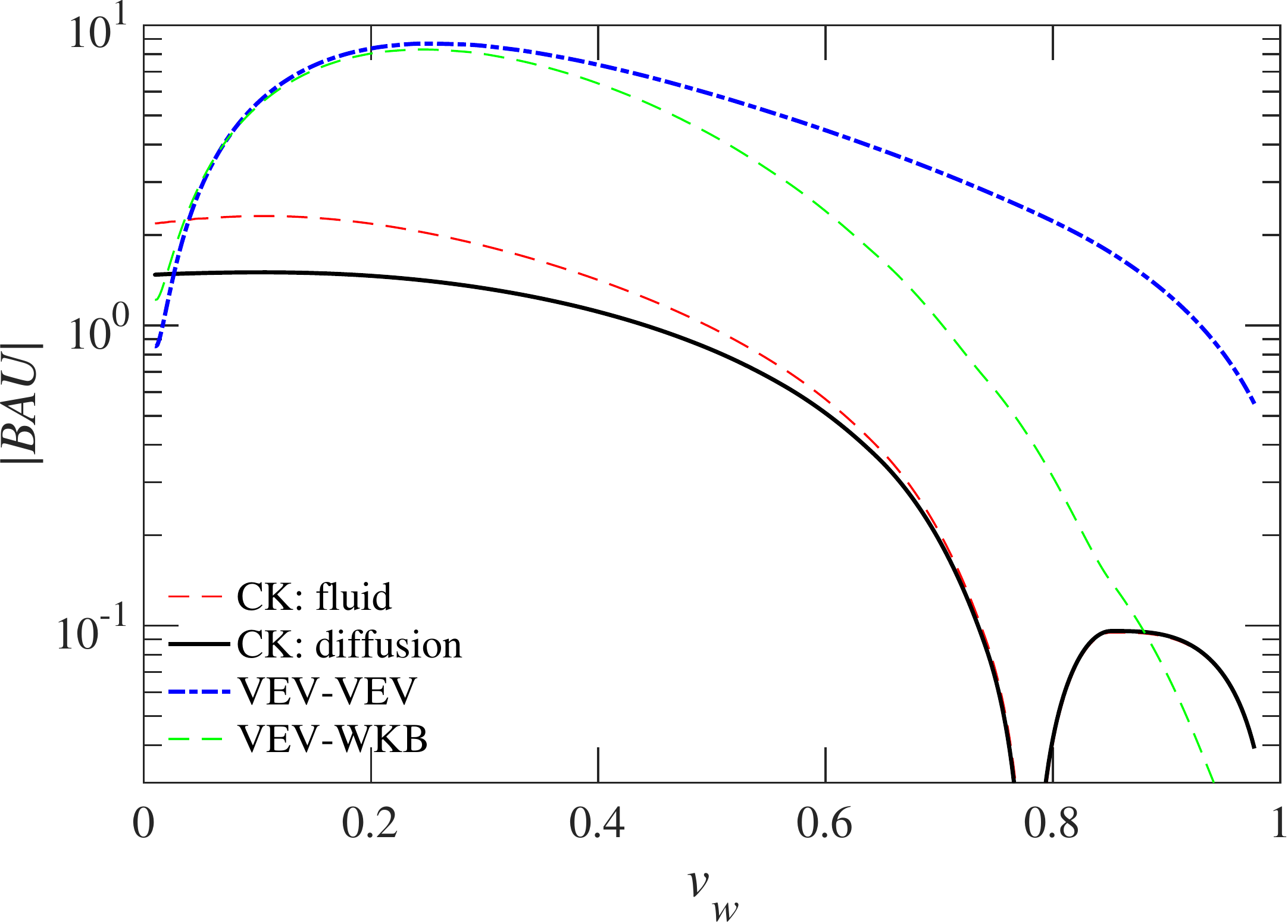}\\ \vskip0.5truecm
\includegraphics[width=0.95\hsize]{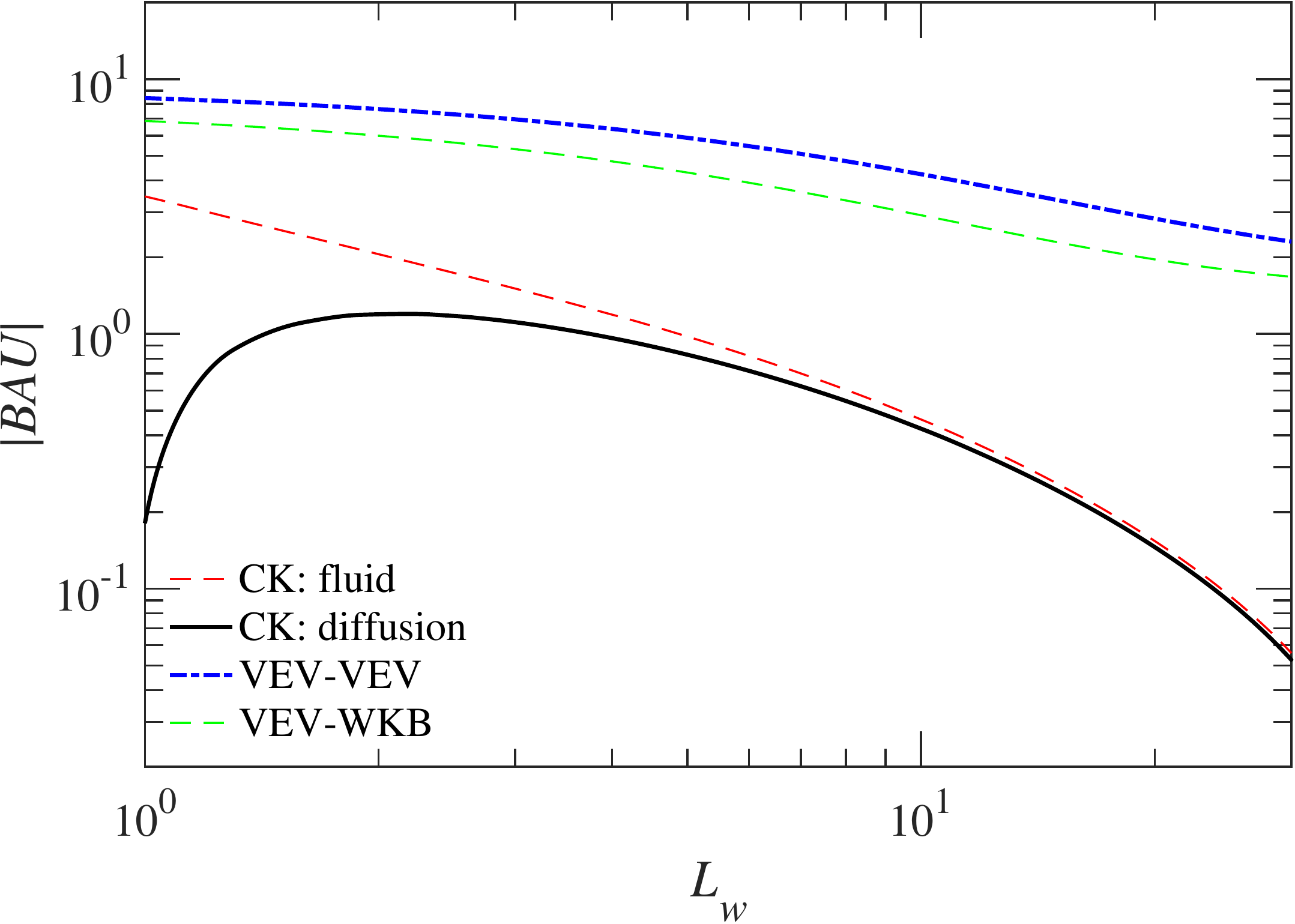} 
\caption{Predicted absolute value of the baryon asymmetry (in units of observed asymmetry)
for the fiducial model as a function of wall velocity $v_w$ (top) and Higgs wall width $L_w$ (bottom).  ``CK-fluid'' denotes the two-moment result derived previously; ``CK-diffusion'' is the diffusion equation approximation (\ref{HNeq}) to the two-moment network; ``VEV-VEV'' is the result from the full VEV-insertion formalism;
``VEV-WKB'' is a hybrid result where the diffusion and collision terms are the same as in the semiclassical (CK) formalism, and only the VEV-insertion source term is different.}
\label{dependences_2}
\end{figure}

We can now compare the semiclassical and VEV-insertion formalism predictions for the BAU on a level playing ground, using the diffusion equations~\eqref{HNeq}. The results are shown in figs.~\ref{dependences_2}. The upper panel displays the absolute value of the BAU $\equiv |\eta_B/\eta_{B,\rm obs}|$ as a function of the wall velocity for our fiducial model~\eqref{eq:fiducial_model} and for the spin-$h$ source. The thick solid black line corresponds to the solution of the diffusion equation~\eqref{HNeq} with the WKB-variables~\eqref{source_conv}. The dashed red line shows for comparison the result of our earlier calculation with improved fluid equations. {The two semiclassical approximations agree remarkably well, in particular for large $v_w$.}

Considering the VEV-insertion formalism, the thick dash-dotted blue line (labeled ``VEV-VEV'') displays the result of using the  parameters~(\ref{Steq}-\ref{eq:C_vev}) in eq.~\eqref{HNeq}. The VEV-insertion method predicts $10-50$ times larger asymmetry than does the semiclassical method.  The thin dashed green line (labeled ``VEV-WKB'') shows the result of using the VEV-insertion source, but WKB diffusion constant and collision terms from eq.~\eqref{source_conv}. {For small $v_w$ the difference between the VEV-insertion method and the WKB-method is dominated by the source terms. However, for $v_w\gsim 0.4$ a further deviation is caused by the $v_w$-dependence of the evolution equations.}

The lower panel displays the dependence of BAU on the wall width $L_w$ for $v_w=0.5$. 
For $L_wT\gsim 2$ the two semiclassical approximations agree very well. The difference at small $L_w$ is expected, because the diffusion approximation neglects several $m^{2\prime}$-corrections included in the fluid equations. As the wall width increases, the semiclassical BAU decreases rapidly. This is the expected behaviour as the system becomes increasingly classical. The VEV-insertion predicts a much slower decrease of BAU with $L_w$.

In our opinion, the semiclassical formalism is the more
reliable one. First, it has a clearly established small expansion parameter: $(L_wT)^{-1}$, and the results show the expected behaviour as a function of this parameter. Second, it has been derived both using WKB-methods~\cite{Cline:1997vk,Cline:2000nw,Cline:2001rk} and from the fundamental 
CTP-formalism~\cite{Kainulainen:2001cn,Kainulainen:2002th,Prokopec:2003pj,Prokopec:2004ic}. Third, the semiclassical limit has been recently shown to arise in a full quantum mechanical treatment without any gradient expansion~\cite{Jukkala:2019slc} (for earlier related work see~\cite{Prokopec:2013ax} and~\cite{Fidler:2011yq,Herranen:2008hi,Herranen:2008hu,Herranen:2008di,Herranen:2009zi,Herranen:2009xi,Herranen:2010mh}.). 

{In the VEV-insertion approach the expansion parameter is not as clearly defined: in addition to expanding in powers of $m(z)/T$, there is also a gradient expansion, which however always results in fewer derivatives acting on the source term than in the semiclassical approach, leading to the different dependence on the wall width shown in fig.~\ref{dependences_2}. Moreover, the source term is infrared singular, cured only by introducing a damping term that is not related to the quantum physics of the problem. This can be seen in eq.\ (\ref{vevsource}), which blows up when the damping rate $\gamma\to 0$.} The lack of convergence of the expansion in $m/T$ was recently investigated in ref.~\cite{Postma:2019scv}, where it was shown that higher order terms are only small if $m/T<\sqrt{\alpha}$, where $\alpha$ is the relevant interaction strength of the fermion of mass $m$. {There is a claim in the literature that the $m/T$ expansion can be avoided \cite{Carena:2000id}, but this is based upon a phenomenological approach which, although superficially similar to the VEV-insertion formalism, does not derive the transport equations from first principles within the CTP formalism.}

%
\section{Conclusions}
\label{sec:dictionary}
%

In this paper we generalized the semiclassical fluid equations for electroweak baryogenesis to the regime of arbitrarily large wall velocities, showing that diffusion remains relatively efficient for $v_w$ exceeding the sound speed in the plasma. As a result,  EWBG can be effective even for very strong transitions corresponding to detonations. We performed a detailed comparison between a new improved network of fluid equations and the previous formulation by Fromme and Huber~\cite{Fromme:2006wx}. For small wall velocities $v_w\lsim 0.1$ the two formulations agree reasonably well, but for larger values, in particular for $v_w$ exceeding the sound speed, the improved formalism is indispensible. 

We then quantitatively compared the semiclassical fluid equations to a competing framework, the VEV insertion method. To do so required reducing the semiclassical fluid equations, consisting of coupled equations for chemical potentials and velocity perturbations, to a set of (WKB) diffusion equations for the particle densities alone (while retaining the full velocity dependence). This was necessary because the VEV-insertion formalism is derived at the level of particle densities, and does not lend itself to a more accurate approximation of the Boltzmann equation in terms of coupled moments (including the velocity perturbation).

The WKB diffusion equations agree very well with the improved fluid equations in the semiclassial picture. The VEV insertion method on the other hand predicts the BAU is a factor of 10-50 times larger than in the semiclassical method. This difference arises mostly (especially for small $v_w$) from the different source terms in the semiclassical and the VEV-insertion schemes.  We argued that the semiclassical results are more reliable, as they have been derived and verified in various different approaches and they, unlike the VEV-insertion results, show the expected parametric behaviours.

Finally, we caution that while our improved fluid equations cure the incorrect dependence predicted by the FH06 network at large $v_w$, they still correspond to a low-order expansion in moments of the
particle distribution functions $f$. One 
might reasonably expect that perturbations  $\delta f$ could be highly nonGaussian, such that going beyond second order in the moment expansion could significantly modify the results presented here. This is a question that deserves further study.

%
\acknowledgments
%

We thank S.\ Lee, M.\ Postma and S.\ Tulin for useful correspondence.
This work was supported by the Academy of Finland grant 318319 and 
NSERC (Natural Sciences and Engineering Research Council, Canada).

%
\begin{appendix}
%

%
\section{Explicit forms for coefficient functions}
\label{sec:app_explicit_forms}
%

All coefficient functions are expressed as integrals over the
distribution functions in the wall rest frame, that in general depend
upon the local particle masses $m(z)/T$ and $v_w$, and they can be
fermionic or bosonic even though the normalization factors $N_i$ are
taken to refer to massless fermions. When evaluating these functions
it is convenient to Lorentz-transform the integration variables:
\bea
E   &=& \gamma_w (E_v - v_wp_{zv}) \nn\\
p_z &=& \gamma_w (p_{zv} - v_wE_v),
\label{Evpv}
\eea
where $E_v$ and $p_{xv}$ are the variables in the plasma
frame. 
One can then use the fact that ${\rm d}^{3}p/E$
is invariant and $\gamma_w(E+ v_w p_z)\to E_v$ so that $f_{0w}\to f_0$
(and similarly for the derivatives of $f_{0w}$). All coefficient
functions can then be written as a two-dimensional integral of a
generic form
\be
\Big\langle \frac{p_z^n}{E^m} V {\cal F}_{0w}\Big\rangle 
= T^{n-m-k} K({\cal F}_0;V;n,m)
\ee
where $k\!=\!0$ for ${\cal F}_{0w} = f_{0w}$, $k\!=\!1$ for ${\cal F}_{0w}=f^\prime_{0w}$ and $k\!=\!2$ for ${\cal F}_{0w}=f^{\prime\prime}_{0w}$ and the dimensionless integral
\bea
K({\cal F}_0;V;n,m) &\equiv& -\frac{3}{\pi^2\gamma_w}\int_x^\infty 
{\rm d}w \int_{-1}^1 {\rm d}y \times \phantom{Hanna}\nn \\
&&\times \,\frac{\tilde p_w\tilde p_z^n}{\tilde E^{m-1}}\,
 V(w,y,v_w,x)\,{\cal F}_0(w),
\phantom{Ha}
\label{eq:Kfun}
\eea
where $\tilde p_w \equiv \sqrt{w^2\!-\!x^2}$ and $\tilde p_z = \gamma_w(y\tilde p_w-wv_w)$ and $\tilde E = \gamma_w(w-v_wy\tilde p_w)$. For $D_\ell$, $Q_\ell$, $Q^e_\ell$ and for $K_1$ the auxiliary function $V \!=\! 1$, while for the CP-odd source functions $Q^{8o}_\ell$ and $Q^{9o}_\ell$ a more complicated structure $V = s_{\rm p}p_z/{E_z}$ appears. For the spin $s$ eigenstates this means:
\be
  V = V_s = \frac{|p_z|}{E_z} = \frac{|\tilde p_z|}{\sqrt{\tilde p_z^2+x^2}}
\ee
and for helicity eigenstates:
\be
  V = V_h \equiv V_s^2 \left(1 - \frac{\tilde p_z^2}{\tilde E^2} \right)^{\!-1/2}.
\ee
where $\tilde p_z$ and $\tilde E$ are as given below~\eqref{eq:Kfun}. Explicitly then:
\bea
  TD_\ell &=&  K(f'_0;1;\ell,\ell ) \nn\\
  T^3Q_\ell &=&  K(f''_0;1;\ell-1,\ell ) \nn\\
  T^2Q^e_\ell &=&  K(f'_0;1;\ell-1,\ell ) \nn\\
  T^3Q^{8o}_\ell &=& \sfrac{1}{2} K(f'_0;V_x;\ell-2,\ell ) \nn\\
  T^5Q^{9o}_\ell &=& \sfrac{1}{4} \big[ K(f'_0;V_x;\ell-2,\ell+2) \nn\\
            && \hskip 0.2truecm-\gamma_wK(f_0'';V_x;\ell-2,\ell+1)\big].
\eea

In particular one can show that some of the coefficient functions are independent of the wall velocity, or that their $v_w$-dependence factorizes simply:
\bea 
  D_0(x)       &=& \frac{1}{\hat N_1} \int {\rm d}^{3}p_v\, f'_0\nn\\
  D_1(x,v_w)   &=& -v_w D_0(x) \nn\\
  Q_1(x,v_w)   &=& \frac{1}{\gamma_w\hat N_1} \int {{\rm d}^{3}p_v\over 2E_v}\,f''_0 \nn\\
  Q^e_1(x,v_w) &=& \frac{1}{\gamma_w\hat N_1} \int {{\rm d}^{3}p_v\over 2E_v}\,f'_0 \nn\\
  K_0(x)       &=& \frac{1}{\hat N_1} \int {\rm d}^{3}p_v\, f_0.
\label{Kfuns}
\eea
All these integrals can be easily reduced to one-dimensional integrals over $E_v$. The function $\bar R$ is a special case, whose one-dimensional integral representation was already given in~\eqref{eq:barR}.

%
\section{VEV-insertion source}
\label{appB}
%

The exact form of the VEV-insertion source has never been derived for the model of
CP-violation (\ref{mteq}) adopted in this work, but similar expressions have been worked 
out for two-Higgs doublet models (2HDMs) where an analogous source is present.  In 2HDMs,
there is an extra suppression factor $\sin^2(2\beta)$ where $\tan\beta = H_2/H_1$ that is
not present in our model.  The source term is therefore similar to eq.~(34) of ref~\cite{Blum:2010by}, except for some typos and an error in that equation \cite{tulin}.  The correct expression is
\bea
\label{vevsource}
	S_{\rm VEV} &=& {v_w N_c m_t(z)^2 \theta' \over 2\pi^2} \int {\rm d}k\,k^2\\
&\times& \; \textrm{Im}\!\left[ \, Z_{t_L}^{p} Z_{t_R}^{h}
  \frac{ n_F(\mathcal{E}^{h*}_{t_R}) - n_F(\mathcal{E}^p_{t_L}) }{ (
    \mathcal{E}_{t_L}^p - \mathcal{E}^{h*}_{t_R} )^2 }\right.\nn\\ &+& \left.Z_{t_L}^{p}
  Z_{t_R}^{p} \frac{ 0 + n_F(\mathcal{E}^p_{ t_R}) +
    n_F(\mathcal{E}^p_{t_L}) }{ ( \mathcal{E}_{t_L}^p +
    \mathcal{E}_{t_R}^p )^2 } \, + \; ( p \leftrightarrow h) \;
\right] \; . \nn
\eea
where the superscripts $p,h$ refer to quasiparticle and hole excitations in the hot plasma,
taken to be in the electroweak symmetric phase and $n_F$ denotes the Fermi-Dirac distribution function evaluated at a complex energy ${\cal E} = E + i\gamma$, where $\gamma$ is the damping rate (thermal width) of the left- or right-handed top quark.  

The error {in the original expression for~\eqref{vevsource}} was that the term ``0'' was originally ``1'', which leads to a UV-divergent integral. It is argued \cite{lee} that normal ordering removes this term, although no derivation has ever been published. The correction to a similar source term was mentioned in ref.\ \cite{Liu:2011jh} (see [31] of that paper). 

The VEV-insertion source has the peculiar property of being singular if the energies ${\cal E}$ are real. The regulating damping rate is dominated by the QCD contribution \cite{Braaten:1992gd}:
\be
  \gamma = \frac{5.7\, g_s^2}{12\pi}\,T . 
\ee
The real parts of the energies are given approximately by
\bea
	E^p &=&  k^2 + {m^2\over m+k},\nn\\
	E^h &=& k\,(1 - 0.45\, e^{-1.5(k/m)^2})
\eea
(these are good analytic fits to the numerical solutions for the poles of the thermally
corrected propagators) with $m$ being the thermal mass for the chirality of interest
\cite{Weldon:1982bn},
\bea
	m_L^2 &=& \left(g_s^2/6 + 3 g_2^2/32 + y_t^2/16\right) T^2 \nn\\
	m_R^2 &=& \left(g_s^2/6 + y_t^2/8\right)
\eea
The wave-function normalization factors are given by
\be
	Z = {E^2 - k^2\over m^2}
\ee
for each kind of particle or hole.  Taking the known values of the coupling constants,
the integral in (\ref{vevsource}) then becomes $I \cong 0.4\,T^3$.  It is normalized in 
this form to be a source for the diffusion equation for particle densities.  To convert
it to a source for the top quark chemical potential, we use the relation $\delta n = 
N_c g_t D_0 T^2 \mu/6$ for a chiral quark with $g_t=2$ spin degrees of freedom.  Hence
$S_t = S_{\rm VEV}/(D_0 T^2)$ to obtain eq.\ (\ref{Steq}).

\end{appendix}
	
%
\bibliographystyle{utphys}
\bibliography{transport6}
%
%
\end{document}